\documentclass[12pt]{article}        
\usepackage{amstex}
\usepackage{amssymb}

\usepackage{epsfig}

\begin{document}                     

\allowdisplaybreaks

\begin{titlepage}


\begin{flushright}
UTHEP-97-0103\\
April,~1997
\end{flushright}

\vspace{1mm}
\begin{center}
{\Large\bf
$e^+$ (Anti)Quark Scattering
 in the Presence\\ of
 the Anomalous HERA Positron-Jet Event Phenomenon
}
\end{center}
\vspace{3mm}

\begin{center}
{\bf S. Jadach$^{a,b}$},
{\bf W. P\l{a}czek$^{c,\star}$},\\
{\em and}\\ 
{\bf B.F.L. Ward$^{a,c,d}$}\\

{\em $^a$CERN, Theory Division, CH-1211 Geneva 23, Switzerland,}\\
{\em $^b$Institute of Nuclear Physics,
  ul. Kawiory 26a, Krak\'ow, Poland,}\\
{\em $^c$Department of Physics and Astronomy,\\
  The University of Tennessee, Knoxville, Tennessee 37996-1200,\\
  $^d$SLAC, Stanford University, Stanford, California 94309} 
\end{center}

\vspace{10mm}
\begin{abstract}
We discuss the leptoquark interpretation of the
anomalous HERA positron-jet events
in the context of the YFS exponentiated Monte Carlo event
generator treatment of the attendant multiple photon radiative
effects for both the would-be signal and the SM background,
wherein finite-$p_T$ photon effects are properly taken into account
and wherein infrared singularities are cancelled to all orders in
$\alpha$. We show that the H1 and ZEUS data are consistent with
such an interpretation for a leptoquark coupling $\lesssim 0.3g_W$,
mass $\sim 200$ GeV, and width $\lesssim 2$GeV. 
Possible future tests are proposed.
\end{abstract}
 
\begin{center}
{\it To be submitted to Phys. Rev. Lett. }
\end{center}

\vspace{1mm}
\begin{flushleft}
{\bf UTHEP-97-0103\\
     April,~1997}
\end{flushleft}

\renewcommand{\baselinestretch}{0.1}
\footnoterule
\noindent
{\footnotesize
\begin{itemize}
\item[${\dagger}$]
Work supported in part by 
the US DoE contract DE-FG05-91ER40627 and DE-AC03-76SF00515,
Polish Government grants KBN 2P03B17210,
the European Union under contract No. ERB-CIPD-CT94-0016,
and Polish-French Collaboration within IN2P3.

\item[${}^{\star}$]
On leave of absence from Institute of Computer Science, Jagiellonian
University, Krak\'ow, Poland.
\end{itemize}
}
 
 
\end{titlepage}

Recently, the H1 and ZEUS Collaborations have reported~\cite{h1,zeus}
an anomalous amount of high $Q^2$ , high $x_{Bj}$ events in the
deep inelastic $e^+$p scattering at HERA. In this paper, we investigate
in detail the interplay between the rigorous treatment of the
attendant multiple photon radiative effects, 
which we treat by the YFS~\cite{yfs} exponentiated Monte Carlo technique
introduced in Ref.~\cite{sjw:1988} and applied to deep-inelastic
lepton-nucleon scattering in Refs.~\cite{lesko-yfs:1992,placzekPHD:1993},
see also Ref.~\cite{zeuthen:1992}, in the event generator LESKO-YFS, and the 
HERA phenomenon if it is interpreted as the exchange of a spin 1 or 0 object
in the s-channel in the reduced 
$e^+{}^{\tiny(}\bar q^{\tiny)}$
parton level process at HERA. For, to our knowledge so far, only
structure function based, zero-$p_T$, treatments of the higher order
QED radiative effects have been applied to the HERA data~\cite{h1,zeus}
so that our work will actually serve two purposes: (1), it will
show that the proper treatment of the higher order finite-$p_T$ photon effects
in the multiple photon radiation cannot account for the phenomenon;
and that, (2), when a resonance of the appropriate coupling
and mass is introduced, the H1 and ZEUS data, in the presence of
the realistic $n(\gamma)$ radiation do indeed agree with the
theoretical expectations. Of course, other possible interpretations
of the data are also possible and we refer the reader to 
Refs.~\cite{guido:1997,pete:1997,dreiner:1997} for further discussion of such 
approaches to the HERA phenomenon.\par

For definiteness, we shall call this object a `leptoquark',
since according to Ref.~\cite{H1-sem}, it has a mass $\sim 200$ GeV
and a width $\lesssim 2.2$ GeV. We stress as it has been already done in
Ref.~\cite{buchmuller:1987} that as long as the couplings of this
object, which may be composite or elementary, are sufficiently
chiral and diagonal in flavor and zero on diquark fields, 
it is not in contradiction with any known physical requirement.
In Ref.~\cite{sjbwzw:1997}, we have discussed the constraints
implied by recent LEP2 data~\cite{jambore97} on this interpretation
of the HERA data due to crossing it into the $e^+e^-\rightarrow hadrons$
channel, as well as the constraints derived in 
Refs.~\cite{leurer:1994a,leurer:1994b}.
See also Refs.~\cite{pete:1997,godfrey:1997,dreiner:1997} for related
and similar analyses of the crossed reaction and see 
Refs.~\cite{guido:1997,blumlein:1997,dreiner:1997} 
for a discussion of contraints
on this interpretation implied by recent TEVATRON data. The net
result of these discussions is that the leptoquark
interpretation is still viable if the coupling strength to 
fermions is small enough, $\lesssim 0.3g_W$, where $g_W$ is the
EW coupling constant. 
Thus, we proceed entirely phenomenologically and try to answer the
very definite question as to whether the HERA data are consistent
with a `leptoquark' explanation if we treat the attendant
higher order radiative effects via the YFS exponentiated Monte Carlo
methods of two of us (S.J. and B.F.L.W.) as they are realized in
the event generator LESKO-YFS~\cite{lesko-yfs:1992,placzekPHD:1993},
for both the signal and the background.\par

More specifically, we first record the 
differential cross section for
$e^+{}^{\tiny(}\bar q^{\tiny)}\rightarrow e^+{}^{\tiny(}\bar q^{\tiny)}$ 
in the presence of the HERA leptoquark
for two models of leptoquark charges. We have, for the outgoing $e^+$
elemental solid angle $d\Omega_{e^+}$ in the CM system,
\begin{equation}
{d\sigma\over d\Omega_{e^+}}= {1\over 64\pi^2\hat{s}}|\bar {\cal M}|^2,
\label{sig}
\end{equation}
where the spin averaged squared matrix element, for the
leptoquark (X) charges $({Q},{Q-1})$ with $Q=5/3$
coupling to left(right)-handed quarks \{model (1)(model (2))\}
and with $Q=4/3$ coupling to left(right)-handed quarks
\{model (3)(model (4))\}, is
\begin{equation}
\label{matsq1L}
\begin{split}
&|\bar {\cal M}|^2_{ \{ {Q}, {Q-1}  \}_H } = \\
&\:\:\:\:\,\left| {(-{1\over2}\delta_{S,0}\delta_{Q,{4\over 3}}+\delta_{S,1}\delta_{Q,{5\over 3}})g_X^2(1-h(H))\over 4D_X(\hat{s})}+\sum_{\rho=A,Z} {G_\rho^2
(v_\rho(q)+a_\rho(q))(v_\rho(e)+a_\rho(e))\over D_\rho(\hat{t})}\right|^2\hat{u}_Q^2
\\&
+ \left| {(-{1\over2}\delta_{S,0}\delta_{Q,{4\over 3}}+\delta_{S,1}\delta_{Q,{5\over 3}})g_X^2(1+h(H))\over 4D_X(\hat{s})}+\sum_{\rho=A,Z} {G_\rho^2
(v_\rho(q)-a_\rho(q))(v_\rho(e)-a_\rho(e))\over D_\rho(\hat{t})}\right|^2\hat{u}_Q^2
\\
&+ \left| {({1\over2}\delta_{S,0}\delta_{Q,{5\over 3}}-\delta_{S,1}\delta_{Q,{4\over 3}})g_X^2(1+h(H))\over 4D_X(\hat{s})}+\sum_{\rho=A,Z} {G_\rho^2
(v_\rho(q)-a_\rho(q))(v_\rho(e)+a_\rho(e))\over D_\rho(\hat{t})}\right|^2\hat{s}_Q^2
\\
&+ \left|{({1\over2}\delta_{S,0}\delta_{Q,{5\over 3}}-\delta_{S,1}\delta_{Q,{4\over 3}})g_X^2(1-h(H))\over 4D_X(\hat{s})}+ \sum_{\rho=A,Z} {G_\rho^2
(v_\rho(q)+a_\rho(q))(v_\rho(e)-a_\rho(e))\over D_\rho(\hat{t})}\right|^2\hat{s}_Q^2,
\end{split}
\end{equation}
where $h(H)=-1(+1)$ for the left ($H\equiv L$)-, right ($H\equiv R$)-handed 
cases respectively, $S$ is the spin of X,
$\delta_{ab}$ is the Kronecker delta function,
$\hat{u}_Q=\delta_{Q,{5\over3}}\hat u+\delta_{Q,{4\over3}}\hat s$,
$\hat{s}_Q=\delta_{Q,{5\over3}}\hat s+\delta_{Q,{4\over 3}}\hat u$ and
where we have defined the following kinematical and dynamical variables:
$ D_X(\hat{s}) = \hat{s}-M_X^2+i\Gamma_X\hat{s}/M_X, \quad M_X= 200~GeV,~\Gamma_X\lesssim 2~GeV$,~
$ D_Z(\hat{t}) = \hat{t}-M_Z^2,\quad
  D_A(\hat{t}) = \hat{t}$,
$ v_Z(f) = {1\over2}I_3-Q_f\sin^2\theta_W,\quad
  a_Z(f) = {1\over2}I_3,$
$  v_A(f) =Q_f,\quad
   a_A(f) =0,$
$ G_Z = e/(\sin\theta_W\cos\theta_W),\quad
  G_A = e,$
$ g_X = (e/\sin\theta_W)(1 + \delta),$
$   \hat{s} = (p_1+q_1)^2,$ ~$\hat{t}=(p_1-p_2)^2$, ~$\hat{u}=(p_1-q_2)^2$,
where $\{p_1,q_1\}$ are the incoming $\{e^+,{}^{\tiny(}\bar q^{\tiny)}\}$
4-momenta respectively
and $\{p_2,q_2\}$ are the outgoing $\{e^+,{}^{\tiny(}\bar q^{\tiny)}\}$ 
4-momenta respectively.
Here, $\delta$ is unknown and is to be varied to see what the
data will allow. (According to Refs.~\cite{sjbwzw:1997,godfrey:1997},
from the crossed reaction constraints we hope to find that
$-1.5\lesssim \delta \lesssim -0.5$ for spin 1.)
$I_3$ is the usual weak isospin 3-component for fermion $f$
and $Q_f$ is its electric charge in units of the positron charge $e$.
We have thus complied with the constraint 
from Refs.~\cite{buchmuller:1987,leurer:1994a,leurer:1994b} that
only quarks of a specific chirality should couple to any
particular leptoquark.
\par

The formulae presented above we have implemented into 
the LESKO-YFS Monte Carlo program~\cite{lesko-yfs:1992,placzekPHD:1993}.
We have performed the technical tests of the matrix elements as implemented
in LESKO-YFS and compared them with analytical results on several approximated
forms of the above matrix elements,  e.g. for $Z+\gamma$ or only 
$Z$ exchange excluded.  
Agreement of 4-5 digits was always found.  
Later, QED corrections were extended (to our 
X-exchange, i.e. non t-channel $Z$, $\gamma$
interaction) 
according to the prescription explained in
Ref.~\cite{colas:1990} for the similar type of modification of
the Monte Carlo KORALZ~\cite{koralz:1994} in Ref.~\cite{sjbwzw:1997}. 

We stress that
the leptoquark-quark-lepton vertices which we have assumed in 
(\ref{matsq1L}), corresponding to the $U_1,\tilde U_1$,
$\tilde V_2,V_2,$ and $U_3$ spin 1 and $R_2,\tilde R_2, S_1, \tilde S_1$
and $S_3$ spin 0 
examples in Ref.~\cite{buchmuller:1987}), 
are intended to be generic
and not exhaustive: it is straightforward to include more general
coupling scenarios into our LESKO-YFS calculational framework, should this
become necessary. We point-out further that
we may identify our states in models 
$(1),\cdots,(4)$ for spin 1,0
respectively with the corresponding charge and mass
eigenstates formed from the 
states $S_L$, $T$,
$\tilde S_R$, $S_R$, $D_L$, $D_R$ and $\tilde D$ in the 
notation of Refs.~\cite{leurer:1994a,leurer:1994b}; for,
at scales $\sim 200$ GeV,
we expect the $SU_{2L}\times U_1$ EW symmetry to be broken with
leptoquark states of the same charge and color mixed into the respective
mass eigenstates and it is these mass eigenstates that we have
used in (\ref{sig}). In other words, from the HERA data we
we may have that $\bar e q$ resonates into the leptoquark $X,~q=u,d$,
models (1) and (2) with t-channel X exchange in $e^+e^-\rightarrow \bar{q}q$
and $F=0$ in the language of Refs.~\cite{buchmuller:1987,leurer:1994a,leurer:1994b},
or that $\bar e \bar q$ resonates 
into the leptoquark $X,~q=u,d$, models (3) and
(4) with u-channel X exchange in $e^+e^-\rightarrow \bar{q}q$ and
$F=-2$ in the language of Refs.~\cite{buchmuller:1987,leurer:1994a,leurer:1994b},
where F is the fermion number of X. 
For definiteness, we have assumed strong isospin symmetry for
simplicity; it is trivial to relax this last assumption, should more data
render this necessary.
For completeness, we then note that the interaction Lagrangian densities
which we used to arrive at the results (\ref{matsq1L}) are, for $S=1$,
${\cal L}_{int,1,H}^{\{5/3,2/3\}} = -(g_X/\sqrt 2)\left(X^{(-5/3)\alpha}_\mu
\bar{e}_H\gamma^\mu{u^\alpha_H}+ X^{(-2/3)\alpha}_\mu
\bar{e}_H\gamma^\mu{d^\alpha_H} + h.c.\right)$, and
${\cal L}_{int,1,H}^{\{4/3,1/3\}} = -{g_X\over \sqrt 2}\left(X^{(-4/3)\alpha}_\mu
\bar{d}^\alpha_H\gamma^\mu{e^c_{H'}}+ X^{(-1/3)\alpha}_\mu
\bar{u}^\alpha_H\gamma^\mu{e^c_{H'}} + h.c.\right)$,
where here $H=L,R$, $H'=L$ when $H=R$ and $H'=R$ when $H=L$,
and we have defined $\psi_{L(R)}=P_{L(R)}\psi$ 
as well as $\psi^c=C\bar\psi^T$ for all $\psi$ when $T$ denotes
transposition. The
the chiral projections are here $P_L=(1-\gamma^5)/2$,$P_R=(1+\gamma^5)/2$
and $C$ is the charge conjugation matrix in an appropriate representation.
Thus, ${\cal L}^{\{5/3,2/3\}}_{int,1,L}$ describes the interactions
used to derive the result (\ref{matsq1L}) for $Q=5/3$, $S=1$, 
$H=L,\,\ldots$, and so on. For $S=0$, the analogous formulas follow from
the corresponding replacements vector $X~\times$ Dirac vector current 
$\Leftrightarrow$ scalar $X~\times$ Dirac scalar current. 
Finally, to facilitate contact with Refs.~\cite{leurer:1994a,leurer:1994b},
we note that, if we make the simplest possible assumption about the
leptoquark weak isospin mixing matrix, that is that our spin 1 states
are composed of only those states in Refs.~\cite{leurer:1994a,leurer:1994b}, 
then we may identify 
$\{T^{(-5/3)},(T^{(-2/3)}-S^{-(2/3)}_L)/\sqrt{2}\}
     \Leftrightarrow \{X^{(-5/3)},X^{(-2/3)}\}$ 
in model (1),
$\{\tilde S^{(-5/3)},S^{(-2/3)}_R\}
     \Leftrightarrow \{X^{(-5/3)},X^{(-2/3)}\}$ 
in model (2),
$\{D^{(-4/3)}_L,D^{(-1/3)}_L\}
     \Leftrightarrow \{X^{(-4/3)},X^{(-1/3)}\}$ 
in model (3), and
$\{D^{(-4/3)}_R,\tilde D^{(-1/3)}\}
     \Leftrightarrow \{X^{(-4/3)},X^{(-1/3)}\}$ 
in model (4),
with the attendant coupling constant relations
$2g=g_X,~\sqrt{2}g=g_X,~\sqrt{2}g=g_X$, and $\sqrt{2}g=g_X$, respectively
where $g$ is the coupling constant in Refs.~\cite{leurer:1994a,leurer:1994b}
(an analogous transformation holds for the spin 0 case).
\par

In the Fig.~1, we exhibit the comparison of the 
expectations of the models (1)-(2) in comparison
to the SM expectations for several observables available
in the HERA data at low $Q^2$ ($2500<Q^2< 15000$ GeV$^2$) for
a value of $\delta$ in the range allowed~\cite{sjbwzw:1997}
by the LEP2 data.
Models (3) and (4) are simply too close to the SM expectation in this
regime to describe the data for any value of $\delta$ of order
1 in magnitude so that we do not bother to indicate 
these latter predictions explicitly in the comparisons with the data.
Values of $\delta$ several orders of magnitude in size
larger than 1 are already excluded by the lack of observation
of pronounced anomalies in the HERA $e^-$ data for models (3) and
(4)~\cite{guido:1997}. 
Shown are the $M_e$ and $y_e$ distributions in Fig.~1(a) and 1(b),
respectively, for $2500<Q_e^2< 15000$ GeV$^2$ and in Fig.~1(c) the $y_e$
distribution for $100$ GeV $< M_e < 180$ GeV 
in the HERA kinematics in the hard scattering CM system for 
which we have the definitions
$s=(p_p+p_1)^2, \quad
Q_e^2=-\hat{t}$,~ $ x_e=Q_e^2/(2(p_1-p_2)p_p)$, ~$ y_e=Q_e^2/(sx_e)$,
so that $M_e=\sqrt{sx_e}$ is the invariant mass of the 
outgoing positron-jet system-- the subscript $e$ means that the 
observable is computed from
variables on the \=e line~\cite{h1,zeus}. Here, $p_p$ is the incoming proton
4-momentum.
Our \=e kinematics is defined
so that all photons within a cone of
half-angle $5^o$ is included in the definition of the outgoing
\=e four-vector, to give a more realistic simulation of the
definition of the outgoing \=e in the HERA detectors.
We also impose a cut on the initial state photon radiation by requiring
that a fraction of the \=e energy lost to this radiation is $\le 0.1$.
We see that for 
$g_X$ such that $\delta \cong -0.7$ for example the $S=0$ models
are well within the experimental errors at HERA~\cite{h1,zeus} of the SM
expectations. By comparison, the $S=1$ models are disfavored at low $y_e$.

 
\begin{figure}[!ht]
\centering
\setlength{\unitlength}{0.05mm}
\begin{picture}(3200,3100)

\put(300,1750){\begin{picture}( 1200,1200)
\put(0,0){\framebox( 1200,1200){ }}
\multiput(  200.00,0)(  266.67,0){   4}{\line(0,1){25}}
\multiput(   13.33,0)(   26.67,0){  45}{\line(0,1){10}}
\multiput(  200.00,1200)(  266.67,0){   4}{\line(0,-1){25}}
\multiput(   13.33,1200)(   26.67,0){  45}{\line(0,-1){10}}
\put( 200,-25){\makebox(0,0)[t]{ $    100 $}}
\put( 467,-25){\makebox(0,0)[t]{ $    150 $}}
\put( 733,-25){\makebox(0,0)[t]{ $    200 $}}
\put(1000,-25){\makebox(0,0)[t]{ $    250 $}}
\put(1050,-120){\makebox(0,0)[t]{ $   M_e [GeV] $}}
\multiput(0,     .00)(0,  240.00){   6}{\line(1,0){25}}
\multiput(0,   24.00)(0,   24.00){  50}{\line(1,0){10}}
\multiput(1200,     .00)(0,  240.00){   6}{\line(-1,0){25}}
\multiput(1200,   24.00)(0,   24.00){  50}{\line(-1,0){10}}
\put(-25,   0){\makebox(0,0)[r]{ $    0.0 $}}
\put(-25, 240){\makebox(0,0)[r]{ $    0.1 $}}
\put(-25, 480){\makebox(0,0)[r]{ $    0.2 $}}
\put(-25, 720){\makebox(0,0)[r]{ $    0.3 $}}
\put(-25, 960){\makebox(0,0)[r]{ $    0.4 $}}
\put(-25,1200){\makebox(0,0)[r]{ $    0.5 $}}
\put(-25,1080){\makebox(0,0)[r]
              {\large $\frac{d\sigma}{dM_e} [\frac{pb}{GeV}]$}}
\put(600,1270){\makebox(0,0)[c]{\large\bf (a)} }
\put(600,1020){\makebox(0,0)[c]{\small $0.1 < y_e < 0.9 $} }
\put(600,1120){\makebox(0,0)[c]
              {\small $2500\,GeV^2 < Q_e^2 < 15000\,GeV^2 $} } 
\end{picture}}
\put(300,1750){\begin{picture}( 1200,1200)
\thicklines 
\newcommand{\x}[3]{\put(#1,#2){\line(1,0){#3}}}
\newcommand{\y}[3]{\put(#1,#2){\line(0,1){#3}}}
\newcommand{\z}[3]{\put(#1,#2){\line(0,-1){#3}}}
\newcommand{\e}[3]{\put(#1,#2){\line(0,1){#3}}}
\y{   0}{   0}{ 868}\x{   0}{ 868}{ 133}
\e{  67}{  867}{   2}
\y{ 133}{ 868}{  72}\x{ 133}{ 940}{ 134}
\e{ 200}{  939}{   2}
\z{ 267}{ 940}{ 192}\x{ 267}{ 748}{ 133}
\e{ 333}{  747}{   2}
\z{ 400}{ 748}{ 241}\x{ 400}{ 507}{ 133}
\e{ 467}{  507}{   2}
\z{ 533}{ 507}{ 272}\x{ 533}{ 235}{ 134}
\e{ 600}{  235}{   0}
\z{ 667}{ 235}{ 156}\x{ 667}{  79}{ 133}
\e{ 733}{   79}{   0}
\z{ 800}{  79}{  58}\x{ 800}{  21}{ 133}
\e{ 867}{   21}{   0}
\z{ 933}{  21}{  18}\x{ 933}{   3}{ 134}
\e{1000}{    3}{   0}
\z{1067}{   3}{   3}\x{1067}{   0}{ 133}
\e{1133}{    0}{   0}
\end{picture}} 
\put(300,1750){\begin{picture}( 1200,1200)
\newcommand{\R}[2]{\put(#1,#2){\circle*{ 20}}}
\newcommand{\E}[3]{\put(#1,#2){\line(0,1){#3}}}
\R{  67}{ 871}
\E{  67}{  870}{   2}
\R{ 200}{ 946}
\E{ 200}{  945}{   2}
\R{ 333}{ 755}
\E{ 333}{  754}{   2}
\R{ 467}{ 515}
\E{ 467}{  514}{   2}
\R{ 600}{ 244}
\E{ 600}{  244}{   0}
\R{ 733}{  99}
\E{ 733}{   99}{   0}
\R{ 867}{  20}
\E{ 867}{   20}{   0}
\R{1000}{   3}
\E{1000}{    3}{   0}
\R{1133}{   0}
\E{1133}{    0}{   0}
\end{picture}} 
\put(300,1750){\begin{picture}( 1200,1200)
\newcommand{\R}[2]{\put(#1,#2){\circle{ 30}}}
\newcommand{\E}[3]{\put(#1,#2){\line(0,1){#3}}}
\R{  67}{ 869}
\E{  67}{  868}{   2}
\R{ 200}{ 945}
\E{ 200}{  944}{   2}
\R{ 333}{ 758}
\E{ 333}{  757}{   2}
\R{ 467}{ 523}
\E{ 467}{  522}{   2}
\R{ 600}{ 259}
\E{ 600}{  259}{   0}
\R{ 733}{ 156}
\E{ 733}{  155}{   2}
\R{ 867}{  18}
\E{ 867}{   18}{   0}
\R{1000}{   3}
\E{1000}{    3}{   0}
\R{1133}{   0}
\E{1133}{    0}{   0}
\end{picture}} 

\put(1900,1750){\begin{picture}( 1200,1200)
\put(0,0){\framebox( 1200,1200){ }}
\multiput(  150.00,0)(  300.00,0){   4}{\line(0,1){25}}
\multiput(     .00,0)(   30.00,0){  41}{\line(0,1){10}}
\multiput(  150.00,1200)(  300.00,0){   4}{\line(0,-1){25}}
\multiput(     .00,1200)(   30.00,0){  41}{\line(0,-1){10}}
\put( 150,-25){\makebox(0,0)[t]{ $   0.2 $}}
\put( 450,-25){\makebox(0,0)[t]{ $   0.4 $}}
\put( 750,-25){\makebox(0,0)[t]{ $   0.6 $}}
\put(1050,-25){\makebox(0,0)[t]{ $   0.8 $}}
\put(1180,-100){\makebox(0,0)[t]{\large $   y_e $}}
\multiput(0,     .00)(0,  300.00){   5}{\line(1,0){25}}
\multiput(0,   30.00)(0,   30.00){  40}{\line(1,0){10}}
\multiput(1200,     .00)(0,  300.00){   5}{\line(-1,0){25}}
\multiput(1200,   30.00)(0,   30.00){  40}{\line(-1,0){10}}
\put(-25,   0){\makebox(0,0)[r]{ $     0 $}}
\put(-25, 300){\makebox(0,0)[r]{ $    25 $}}
\put(-25, 600){\makebox(0,0)[r]{ $    50 $}}
\put(-25, 900){\makebox(0,0)[r]{ $    75 $}}
\put(-25,1200){\makebox(0,0)[r]{ $   100 $}}
\put(-25,1050){\makebox(0,0)[r]{\large $ \frac{d\sigma}{dy_e}[pb] $}}
\put(600,1270){\makebox(0,0)[c]{\large\bf (b)} }
\put(650,1120){\makebox(0,0)[c]
              {\small $2500\,GeV^2 < Q_e^2 < 15000\,GeV^2 $} } 
\end{picture}}
\put(1900,1750){\begin{picture}( 1200,1200)
\thicklines 
\newcommand{\x}[3]{\put(#1,#2){\line(1,0){#3}}}
\newcommand{\y}[3]{\put(#1,#2){\line(0,1){#3}}}
\newcommand{\z}[3]{\put(#1,#2){\line(0,-1){#3}}}
\newcommand{\e}[3]{\put(#1,#2){\line(0,1){#3}}}
\y{   0}{   0}{ 912}\x{   0}{ 912}{ 150}
\e{  75}{  911}{   2}
\z{ 150}{ 912}{  40}\x{ 150}{ 872}{ 150}
\e{ 225}{  871}{   2}
\z{ 300}{ 872}{ 169}\x{ 300}{ 703}{ 150}
\e{ 375}{  703}{   2}
\z{ 450}{ 703}{ 142}\x{ 450}{ 561}{ 150}
\e{ 525}{  560}{   2}
\z{ 600}{ 561}{ 104}\x{ 600}{ 457}{ 150}
\e{ 675}{  456}{   2}
\z{ 750}{ 457}{  77}\x{ 750}{ 380}{ 150}
\e{ 825}{  380}{   2}
\z{ 900}{ 380}{  76}\x{ 900}{ 304}{ 150}
\e{ 975}{  303}{   2}
\z{1050}{ 304}{ 178}\x{1050}{ 126}{ 150}
\e{1125}{  126}{   0}
\end{picture}} 
\put(1900,1750){\begin{picture}( 1200,1200)
\newcommand{\R}[2]{\put(#1,#2){\circle*{ 20}}}
\newcommand{\E}[3]{\put(#1,#2){\line(0,1){#3}}}
\R{  75}{ 930}
\E{  75}{  929}{   2}
\R{ 225}{ 889}
\E{ 225}{  888}{   2}
\R{ 375}{ 717}
\E{ 375}{  716}{   2}
\R{ 525}{ 566}
\E{ 525}{  566}{   2}
\R{ 675}{ 461}
\E{ 675}{  460}{   2}
\R{ 825}{ 383}
\E{ 825}{  383}{   2}
\R{ 975}{ 306}
\E{ 975}{  306}{   2}
\R{1125}{ 128}
\E{1125}{  127}{   0}
\end{picture}} 
\put(1900,1750){\begin{picture}( 1200,1200)
\newcommand{\R}[2]{\put(#1,#2){\circle{ 30}}}
\newcommand{\E}[3]{\put(#1,#2){\line(0,1){#3}}}
\R{  75}{ 979}
\E{  75}{  978}{   2}
\R{ 225}{ 923}
\E{ 225}{  922}{   2}
\R{ 375}{ 736}
\E{ 375}{  735}{   2}
\R{ 525}{ 568}
\E{ 525}{  567}{   2}
\R{ 675}{ 460}
\E{ 675}{  459}{   2}
\R{ 825}{ 382}
\E{ 825}{  381}{   2}
\R{ 975}{ 304}
\E{ 975}{  304}{   2}
\R{1125}{ 126}
\E{1125}{  126}{   0}
\end{picture}} 
 
\put(1900,200){\begin{picture}( 1200,1200)
\put(0,0){\framebox( 1200,1200){ }}
\multiput(  150.00,0)(  300.00,0){   4}{\line(0,1){25}}
\multiput(     .00,0)(   30.00,0){  41}{\line(0,1){10}}
\multiput(  150.00,1200)(  300.00,0){   4}{\line(0,-1){25}}
\multiput(     .00,1200)(   30.00,0){  41}{\line(0,-1){10}}
\put( 150,-25){\makebox(0,0)[t]{ $   0.2 $}}
\put( 450,-25){\makebox(0,0)[t]{ $   0.4 $}}
\put( 750,-25){\makebox(0,0)[t]{ $   0.6 $}}
\put(1050,-25){\makebox(0,0)[t]{ $   0.8 $}}
\put(1180,-100){\makebox(0,0)[t]{\large $   y_e $}}
\multiput(0,     .00)(0,  600.00){   3}{\line(1,0){25}}
\multiput(0,  180.60)(0,  600.00){   2}{\line(1,0){10}}
\multiput(0,  286.26)(0,  600.00){   2}{\line(1,0){10}}
\multiput(0,  361.26)(0,  600.00){   2}{\line(1,0){10}}
\multiput(0,  419.40)(0,  600.00){   2}{\line(1,0){10}}
\multiput(0,  466.92)(0,  600.00){   2}{\line(1,0){10}}
\multiput(0,  507.06)(0,  600.00){   2}{\line(1,0){10}}
\multiput(0,  541.86)(0,  600.00){   2}{\line(1,0){10}}
\multiput(0,  572.52)(0,  600.00){   2}{\line(1,0){10}}
\multiput(1200,     .00)(0,  600.00){   3}{\line(-1,0){25}}
\multiput(1200,  180.60)(0,  600.00){   2}{\line(-1,0){10}}
\multiput(1200,  286.26)(0,  600.00){   2}{\line(-1,0){10}}
\multiput(1200,  361.26)(0,  600.00){   2}{\line(-1,0){10}}
\multiput(1200,  419.40)(0,  600.00){   2}{\line(-1,0){10}}
\multiput(1200,  466.92)(0,  600.00){   2}{\line(-1,0){10}}
\multiput(1200,  507.06)(0,  600.00){   2}{\line(-1,0){10}}
\multiput(1200,  541.86)(0,  600.00){   2}{\line(-1,0){10}}
\multiput(1200,  572.52)(0,  600.00){   2}{\line(-1,0){10}}
\put(-25,   0){\makebox(0,0)[r]{\large $    1 $}}
\put(-25, 600){\makebox(0,0)[r]{\large $    10 $}}
\put(-25,1200){\makebox(0,0)[r]{\large $   10^2 $}}
\put(-25,1050){\makebox(0,0)[r]{\large $ \frac{d\sigma}{dy_e}[pb] $}}
\put(600,1270){\makebox(0,0)[c]{\large\bf (c)} }
\put(750,1120){\makebox(0,0)[c]{\small $100\,GeV < M_e < 180\,GeV $} } 
\end{picture}}
\put(1900,200){\begin{picture}( 1200,1200)
\thicklines 
\newcommand{\x}[3]{\put(#1,#2){\line(1,0){#3}}}
\newcommand{\y}[3]{\put(#1,#2){\line(0,1){#3}}}
\newcommand{\z}[3]{\put(#1,#2){\line(0,-1){#3}}}
\newcommand{\e}[3]{\put(#1,#2){\line(0,1){#3}}}
\y{   0}{   0}{1086}\x{   0}{1086}{ 150}
\e{  75}{ 1086}{   0}
\z{ 150}{1086}{   1}\x{ 150}{1085}{ 150}
\e{ 225}{ 1084}{   0}
\z{ 300}{1085}{ 183}\x{ 300}{ 902}{ 150}
\e{ 375}{  902}{   0}
\z{ 450}{ 902}{ 171}\x{ 450}{ 731}{ 150}
\e{ 525}{  731}{   2}
\z{ 600}{ 731}{ 143}\x{ 600}{ 588}{ 150}
\e{ 675}{  587}{   2}
\z{ 750}{ 588}{ 126}\x{ 750}{ 462}{ 150}
\e{ 825}{  461}{   2}
\z{ 900}{ 462}{ 108}\x{ 900}{ 354}{ 150}
\e{ 975}{  353}{   2}
\z{1050}{ 354}{  91}\x{1050}{ 263}{ 150}
\e{1125}{  262}{   2}
\end{picture}} 
\put(1900,200){\begin{picture}( 1200,1200)
\newcommand{\R}[2]{\put(#1,#2){\circle*{ 20}}}
\newcommand{\E}[3]{\put(#1,#2){\line(0,1){#3}}}
\R{  75}{1089}
\E{  75}{ 1089}{   0}
\R{ 225}{1087}
\E{ 225}{ 1087}{   0}
\R{ 375}{ 906}
\E{ 375}{  905}{   0}
\R{ 525}{ 736}
\E{ 525}{  735}{   2}
\R{ 675}{ 594}
\E{ 675}{  594}{   2}
\R{ 825}{ 471}
\E{ 825}{  470}{   2}
\R{ 975}{ 366}
\E{ 975}{  364}{   2}
\R{1125}{ 277}
\E{1125}{  276}{   2}
\end{picture}} 
\put(1900,200){\begin{picture}( 1200,1200)
\newcommand{\R}[2]{\put(#1,#2){\circle{ 30}}}
\newcommand{\E}[3]{\put(#1,#2){\line(0,1){#3}}}
\R{  75}{1094}
\E{  75}{ 1093}{   0}
\R{ 225}{1090}
\E{ 225}{ 1090}{   0}
\R{ 375}{ 908}
\E{ 375}{  908}{   0}
\R{ 525}{ 738}
\E{ 525}{  737}{   2}
\R{ 675}{ 594}
\E{ 675}{  593}{   2}
\R{ 825}{ 467}
\E{ 825}{  467}{   2}
\R{ 975}{ 358}
\E{ 975}{  356}{   2}
\R{1125}{ 265}
\E{1125}{  263}{   2}
\end{picture}} 

\put(200,300){\begin{picture}( 1200,1200)
\thicklines
\put(0  ,950){\line(1,0){180} }
\put(220,930){Standard Model}
\thinlines
\put(220,800){Leptoquarks $X^{(-5/3,-2/3)}$:}
\multiput(10 ,680)(50,0){4}{\circle*{20} }
\multiput(10 ,530)(50,0){4}{\circle{30} }
\put(220,650){\it Scalar, $\delta_S = -0.72$ }
\put(220,500){\it Vector, $\delta_V = -0.68$ }
\put(0,350){$M_X = 200\,GeV,\;\Gamma_X = 2\,GeV$}
\end{picture}} %
\end{picture} 
\caption[fig1]{\small\sf 
Distributions of $M_e$ and $y_e$ obtained from LESKO-YFS for the
Standard Model process (solid histograms) and in the presence
of the scalar (full circles) and vector (open circles)
leptoquarks $X^{(-5/3,-2/3)}$ in models (1) and (2) 
-- both models give identical distributions. 
All the results were obtained for the MRSA~\cite{MRSA:1994} parametrization
of the proton structure functions. 
}
\label{fig1}

\end{figure}
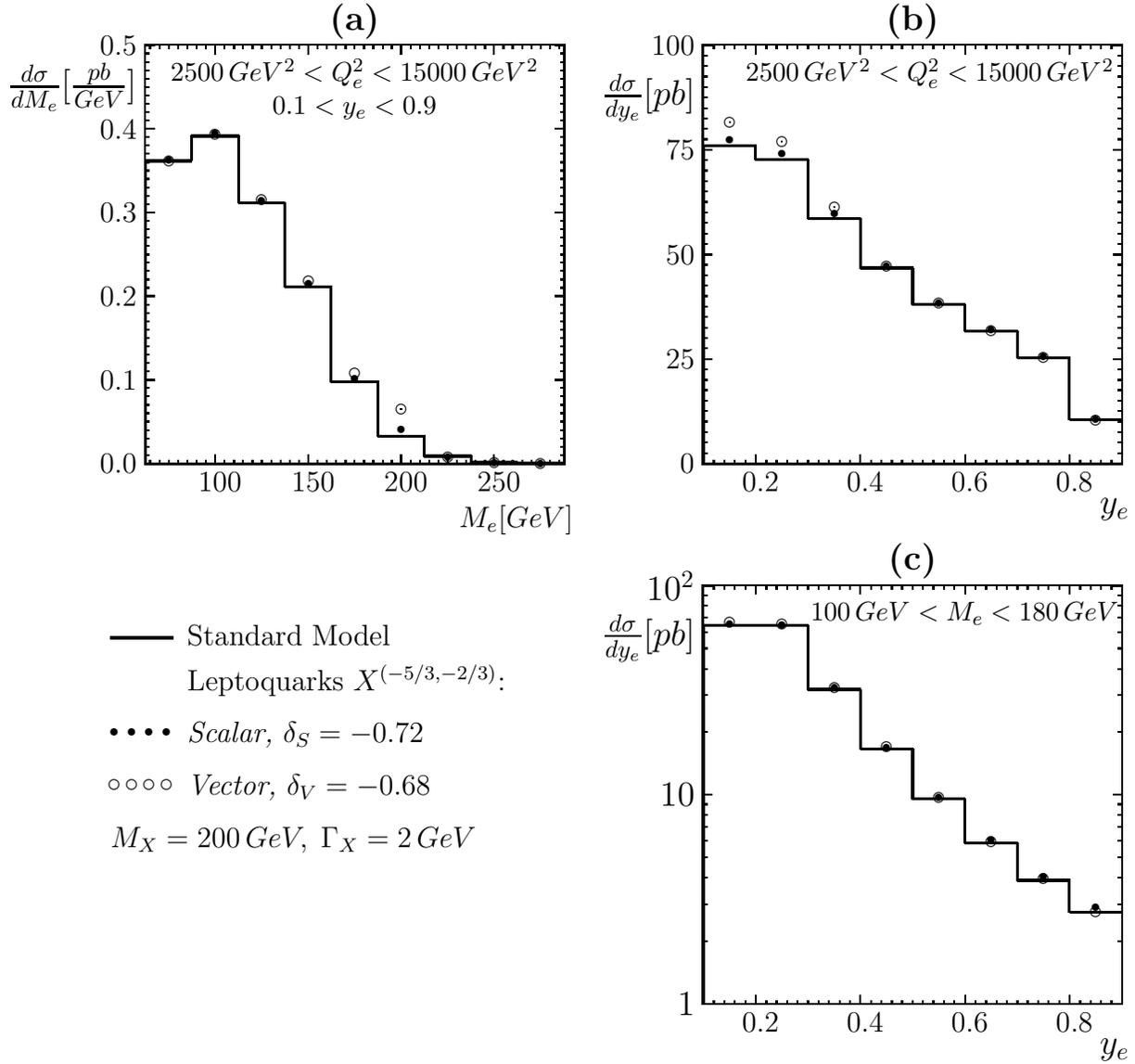 

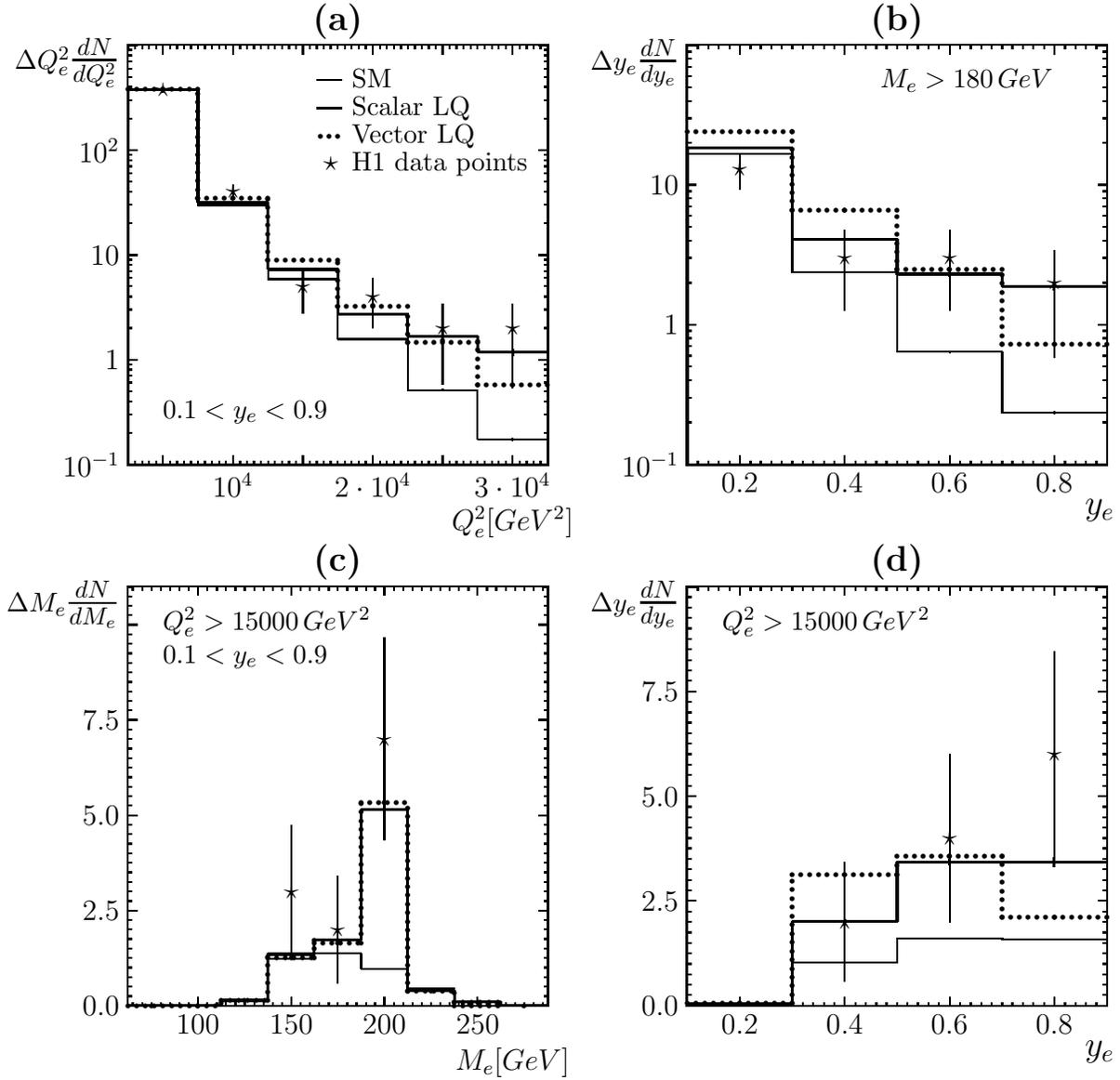
\begin{figure}[!ht]
\centering
\setlength{\unitlength}{0.05mm}
\begin{picture}(3200,3100)

\put(300,1750){\begin{picture}( 1200,1200)
\put(0,0){\framebox( 1200,1200){ }}
\multiput(  100.00,0)(  200.00,0){   6}{\line(0,1){25}}
\multiput(     .00,0)(   20.00,0){  61}{\line(0,1){10}}
\multiput(  100.00,1200)(  200.00,0){   6}{\line(0,-1){25}}
\multiput(     .00,1200)(   20.00,0){  61}{\line(0,-1){10}}
\put( 300,-25){\makebox(0,0)[t]{$     10^{   4} $}}
\put( 700,-25){\makebox(0,0)[t]{$    2\cdot 10^{   4} $}}
\put(1100,-25){\makebox(0,0)[t]{$    3\cdot 10^{   4} $}}
\put(1100,-120){\makebox(0,0)[t]{$Q_e^2 [GeV^2] $}}
\multiput(0,     .00)(0,  300.00){   5}{\line(1,0){25}}
\multiput(0,   90.30)(0,  300.00){   4}{\line(1,0){10}}
\multiput(0,  143.13)(0,  300.00){   4}{\line(1,0){10}}
\multiput(0,  180.63)(0,  300.00){   4}{\line(1,0){10}}
\multiput(0,  209.70)(0,  300.00){   4}{\line(1,0){10}}
\multiput(0,  233.46)(0,  300.00){   4}{\line(1,0){10}}
\multiput(0,  253.53)(0,  300.00){   4}{\line(1,0){10}}
\multiput(0,  270.93)(0,  300.00){   4}{\line(1,0){10}}
\multiput(0,  286.26)(0,  300.00){   4}{\line(1,0){10}}
\multiput(1200,     .00)(0,  300.00){   5}{\line(-1,0){25}}
\multiput(1200,   90.30)(0,  300.00){   4}{\line(-1,0){10}}
\multiput(1200,  143.13)(0,  300.00){   4}{\line(-1,0){10}}
\multiput(1200,  180.63)(0,  300.00){   4}{\line(-1,0){10}}
\multiput(1200,  209.70)(0,  300.00){   4}{\line(-1,0){10}}
\multiput(1200,  233.46)(0,  300.00){   4}{\line(-1,0){10}}
\multiput(1200,  253.53)(0,  300.00){   4}{\line(-1,0){10}}
\multiput(1200,  270.93)(0,  300.00){   4}{\line(-1,0){10}}
\multiput(1200,  286.26)(0,  300.00){   4}{\line(-1,0){10}}
\put(-25,   0){\makebox(0,0)[r]{$   10^{-1} $}}
\put(-25, 300){\makebox(0,0)[r]{$     1 $}}
\put(-25, 600){\makebox(0,0)[r]{$    10 $}}
\put(-25,900){\makebox(0,0)[r]{$    10^{2} $}}
\put(-25,1150){\makebox(0,0)[r]
              {{\small $\Delta Q_e^2$}{\large $\frac{dN}{dQ_e^2}$}}}
\put(600,1270){\makebox(0,0)[c]{\large\bf (a)} }
\put(100, 150){\makebox(0,0)[l]{\small $0.1 < y_e < 0.9$}}
\thinlines
\put(540,1100){\line(1,0){70} }
\put(640,1080){\small SM}
\thicklines
\put(540,1020){\line(1,0){70} }
\put(640,1000){\small Scalar LQ}
\multiput(550,940)(25,0){3}{\circle*{12} }
\put(640, 920){\small Vector LQ}
\put(560, 840){$\star$}
\put(640, 840){\small H1 data points}
\thinlines
\end{picture}}
\put(300,1750){\begin{picture}( 1200,1200)
\newcommand{\R}[2]{\put(#1,#2){\makebox(0,0){$\star$}}}
\newcommand{\E}[3]{\put(#1,#2){\line(0,1){#3}}}
\R{ 100}{1072}
\E{ 100}{ 1065}{  13}
\R{ 300}{ 781}
\E{ 300}{  758}{  42}
\R{ 500}{ 510}
\E{ 500}{  432}{ 125}
\R{ 700}{ 481}
\E{ 700}{  390}{ 143}
\R{ 900}{ 390}
\E{ 900}{  230}{ 230}
\R{1100}{ 390}
\E{1100}{  230}{ 230}
\end{picture}} 
\put(300,1750){\begin{picture}( 1200,1200)
\thinlines 
\newcommand{\x}[3]{\put(#1,#2){\line(1,0){#3}}}
\newcommand{\y}[3]{\put(#1,#2){\line(0,1){#3}}}
\newcommand{\z}[3]{\put(#1,#2){\line(0,-1){#3}}}
\newcommand{\e}[3]{\put(#1,#2){\line(0,1){#3}}}
\y{   0}{   0}{1072}\x{   0}{1072}{ 200}
\e{ 100}{ 1072}{   0}
\z{ 200}{1072}{ 330}\x{ 200}{ 742}{ 200}
\e{ 300}{  742}{   0}
\z{ 400}{ 742}{ 212}\x{ 400}{ 530}{ 200}
\e{ 500}{  530}{   0}
\z{ 600}{ 530}{ 172}\x{ 600}{ 358}{ 200}
\e{ 700}{  357}{   2}
\z{ 800}{ 358}{ 146}\x{ 800}{ 212}{ 200}
\e{ 900}{  210}{   4}
\z{1000}{ 212}{ 140}\x{1000}{  72}{ 200}
\e{1100}{   69}{   6}
\end{picture}} 
\put(300,1750){\begin{picture}( 1200,1200)
\thicklines 
\newcommand{\x}[3]{\put(#1,#2){\line(1,0){#3}}}
\newcommand{\y}[3]{\put(#1,#2){\line(0,1){#3}}}
\newcommand{\z}[3]{\put(#1,#2){\line(0,-1){#3}}}
\newcommand{\e}[3]{\put(#1,#2){\line(0,1){#3}}}
\y{   0}{   0}{1073}\x{   0}{1073}{ 200}
\e{ 100}{ 1073}{   0}
\z{ 200}{1073}{ 323}\x{ 200}{ 750}{ 200}
\e{ 300}{  750}{   0}
\z{ 400}{ 750}{ 191}\x{ 400}{ 559}{ 200}
\e{ 500}{  558}{   2}
\z{ 600}{ 559}{ 128}\x{ 600}{ 431}{ 200}
\e{ 700}{  429}{   4}
\z{ 800}{ 431}{  65}\x{ 800}{ 366}{ 200}
\e{ 900}{  361}{   8}
\z{1000}{ 366}{  45}\x{1000}{ 321}{ 200}
\e{1100}{  312}{  18}
\end{picture}} 
\put(300,1750){\begin{picture}( 1200,1200)
\newcommand{\R}[2]{\put(#1,#2){\circle*{ 12}}}
\newcommand{\E}[3]{\put(#1,#2){\line(0,1){#3}}}
\multiput(  0,1075)(25,0){9}{\circle*{12} }
\R{ 100}{1075}
\E{ 100}{ 1075}{   0}
\multiput(200,762)(0,25){13}{\circle*{12} }
\multiput(200,762)(25,0){9}{\circle*{12} }
\R{ 300}{ 762}
\E{ 300}{  762}{   0}
\multiput(400,585)(0,25){8}{\circle*{12} }
\multiput(400,585)(25,0){9}{\circle*{12} }
\R{ 500}{ 585}
\E{ 500}{  584}{   2}
\multiput(600,453)(0,25){6}{\circle*{12} }
\multiput(600,453)(25,0){9}{\circle*{12} }
\R{ 700}{ 455}
\E{ 700}{  453}{   4}
\multiput(800,349)(0,25){4}{\circle*{12} }
\multiput(800,349)(25,0){9}{\circle*{12} }
\R{ 900}{ 354}
\E{ 900}{  349}{  10}
\multiput(1000,228)(0,25){6}{\circle*{12} }
\multiput(1000,228)(25,0){9}{\circle*{12} }
\R{1100}{ 228}
\E{1100}{  220}{  14}
\end{picture}} 

\put(1900,1750){\begin{picture}( 1200,1200)
\put(0,0){\framebox( 1200,1200){ }}
\multiput(  150.00,0)(  300.00,0){   4}{\line(0,1){25}}
\multiput(     .00,0)(   30.00,0){  41}{\line(0,1){10}}
\multiput(  150.00,1200)(  300.00,0){   4}{\line(0,-1){25}}
\multiput(     .00,1200)(   30.00,0){  41}{\line(0,-1){10}}
\put( 150,-25){\makebox(0,0)[t]{$    0.2 $}}
\put( 450,-25){\makebox(0,0)[t]{$    0.4 $}}
\put( 750,-25){\makebox(0,0)[t]{$    0.6 $}}
\put(1050,-25){\makebox(0,0)[t]{$    0.8 $}}
\put(1180,-100){\makebox(0,0)[t]{\large $    y_e $}}
\multiput(0,     .00)(0,  400.00){   4}{\line(1,0){25}}
\multiput(0,     .00)(0,  400.00){   4}{\line(1,0){25}}
\multiput(0,  120.40)(0,  400.00){   3}{\line(1,0){10}}
\multiput(0,  190.84)(0,  400.00){   3}{\line(1,0){10}}
\multiput(0,  240.84)(0,  400.00){   3}{\line(1,0){10}}
\multiput(0,  279.60)(0,  400.00){   3}{\line(1,0){10}}
\multiput(0,  311.28)(0,  400.00){   3}{\line(1,0){10}}
\multiput(0,  338.04)(0,  400.00){   3}{\line(1,0){10}}
\multiput(0,  361.24)(0,  400.00){   3}{\line(1,0){10}}
\multiput(0,  381.68)(0,  400.00){   3}{\line(1,0){10}}
\multiput(1200,     .00)(0,  400.00){   4}{\line(-1,0){25}}
\multiput(1200,  120.40)(0,  400.00){   3}{\line(-1,0){10}}
\multiput(1200,  190.84)(0,  400.00){   3}{\line(-1,0){10}}
\multiput(1200,  240.84)(0,  400.00){   3}{\line(-1,0){10}}
\multiput(1200,  279.60)(0,  400.00){   3}{\line(-1,0){10}}
\multiput(1200,  311.28)(0,  400.00){   3}{\line(-1,0){10}}
\multiput(1200,  338.04)(0,  400.00){   3}{\line(-1,0){10}}
\multiput(1200,  361.24)(0,  400.00){   3}{\line(-1,0){10}}
\multiput(1200,  381.68)(0,  400.00){   3}{\line(-1,0){10}}
\put(-25,   0){\makebox(0,0)[r]{$   10^{-1} $}}
\put(-25, 400){\makebox(0,0)[r]{$     1 $}}
\put(-25, 800){\makebox(0,0)[r]{$    10 $}}
\put(-25,1150){\makebox(0,0)[r]
              {{\small $\Delta y_e$}{\large $\frac{dN}{dy_e}$}}}
\put(600,1270){\makebox(0,0)[c]{\large\bf (b)} }
\put(800,1100){\makebox(0,0)[c]{\small $M_e>180\,GeV$}}
\end{picture}}
\put(1900,1750){\begin{picture}( 1200,1200)
\newcommand{\R}[2]{\put(#1,#2){\makebox(0,0){$\star$}}}
\newcommand{\E}[3]{\put(#1,#2){\line(0,1){#3}}}
\R{ 150}{ 846}
\E{ 150}{  789}{  98}
\R{ 450}{ 591}
\E{ 450}{  441}{ 229}
\R{ 750}{ 591}
\E{ 750}{  441}{ 229}
\R{1050}{ 520}
\E{1050}{  307}{ 306}
\end{picture}} 
\put(1900,1750){\begin{picture}( 1200,1200)
\thinlines 
\newcommand{\x}[3]{\put(#1,#2){\line(1,0){#3}}}
\newcommand{\y}[3]{\put(#1,#2){\line(0,1){#3}}}
\newcommand{\z}[3]{\put(#1,#2){\line(0,-1){#3}}}
\newcommand{\e}[3]{\put(#1,#2){\line(0,1){#3}}}
\y{   0}{   0}{ 888}\x{   0}{ 888}{ 300}
\e{ 150}{  888}{   0}
\z{ 300}{ 888}{ 338}\x{ 300}{ 550}{ 300}
\e{ 450}{  549}{   2}
\z{ 600}{ 550}{ 228}\x{ 600}{ 322}{ 300}
\e{ 750}{  320}{   4}
\z{ 900}{ 322}{ 174}\x{ 900}{ 148}{ 300}
\e{1050}{  144}{   6}
\end{picture}} 
\put(1900,1750){\begin{picture}( 1200,1200)
\thicklines 
\newcommand{\x}[3]{\put(#1,#2){\line(1,0){#3}}}
\newcommand{\y}[3]{\put(#1,#2){\line(0,1){#3}}}
\newcommand{\z}[3]{\put(#1,#2){\line(0,-1){#3}}}
\newcommand{\e}[3]{\put(#1,#2){\line(0,1){#3}}}
\y{   0}{   0}{ 906}\x{   0}{ 906}{ 300}
\e{ 150}{  906}{   0}
\z{ 300}{ 906}{ 261}\x{ 300}{ 645}{ 300}
\e{ 450}{  643}{   4}
\z{ 600}{ 645}{ 100}\x{ 600}{ 545}{ 300}
\e{ 750}{  540}{  10}
\z{ 900}{ 545}{  36}\x{ 900}{ 509}{ 300}
\e{1050}{  499}{  22}
\end{picture}} 
\put(1900,1750){\begin{picture}( 1200,1200)
\newcommand{\R}[2]{\put(#1,#2){\circle*{ 12}}}
\newcommand{\E}[3]{\put(#1,#2){\line(0,1){#3}}}
\multiput(  0,953)(25,0){13}{\circle*{12} }
\R{ 150}{ 953}
\E{ 150}{  953}{   2}
\multiput(300,728)(0,25){10}{\circle*{12} }
\multiput(300,728)(25,0){13}{\circle*{12} }
\R{ 450}{ 728}
\E{ 450}{  726}{   4}
\multiput(600,558)( 0,25){7}{\circle*{12} }
\multiput(600,558)(25,0){13}{\circle*{12} }
\R{ 750}{ 558}
\E{ 750}{  553}{  10}
\multiput(900,343)( 0,25){9}{\circle*{12} }
\multiput(900,343)(25,0){13}{\circle*{12} }
\R{1050}{ 343}
\E{1050}{  335}{  16}
\end{picture}} 

\put(300,200){\begin{picture}( 1200,1200)
\put(0,0){\framebox( 1200,1200){ }}
\multiput(  200.00,0)(  266.67,0){   4}{\line(0,1){25}}
\multiput(   13.33,0)(   26.67,0){  45}{\line(0,1){10}}
\multiput(  200.00,1200)(  266.67,0){   4}{\line(0,-1){25}}
\multiput(   13.33,1200)(   26.67,0){  45}{\line(0,-1){10}}
\put( 200,-25){\makebox(0,0)[t]{$    100 $}}
\put( 467,-25){\makebox(0,0)[t]{$    150 $}}
\put( 733,-25){\makebox(0,0)[t]{$    200 $}}
\put(1000,-25){\makebox(0,0)[t]{$    250 $}}
\put(1100,-120){\makebox(0,0)[t]{$    M_e [GeV] $}}
\multiput(0,     .00)(0,  272.73){   5}{\line(1,0){25}}
\multiput(0,   27.27)(0,   27.27){  44}{\line(1,0){10}}
\multiput(1200,     .00)(0,  272.73){   5}{\line(-1,0){25}}
\multiput(1200,   27.27)(0,   27.27){  44}{\line(-1,0){10}}
\put(-25,   0){\makebox(0,0)[r]{$    0.0 $}}
\put(-25, 273){\makebox(0,0)[r]{$    2.5 $}}
\put(-25, 545){\makebox(0,0)[r]{$    5.0 $}}
\put(-25, 818){\makebox(0,0)[r]{$    7.5 $}}
\put(-25,1150){\makebox(0,0)[r]
              {{\small $\Delta M_e$}{\large $\frac{dN}{dM_e}$}}}
\put(600,1270){\makebox(0,0)[c]{\large\bf (c)} }
\put(100,1000){\makebox(0,0)[l]{\small $0.1 < y_e < 0.9$}}
\put(100,1100){\makebox(0,0)[l]{\small $Q_e^2 > 15000\,GeV^2$}}
\end{picture}}
\put(300,200){\begin{picture}( 1200,1200)
\newcommand{\R}[2]{\put(#1,#2){\makebox(0,0){$\star$}}}
\newcommand{\E}[3]{\put(#1,#2){\line(0,1){#3}}}
\R{ 467}{ 327}
\E{ 467}{  138}{ 378}
\R{ 600}{ 218}
\E{ 600}{   64}{ 308}
\R{ 733}{ 764}
\E{ 733}{  475}{ 578}
\end{picture}} 
\put(300,200){\begin{picture}( 1200,1200)
\thinlines 
\newcommand{\x}[3]{\put(#1,#2){\line(1,0){#3}}}
\newcommand{\y}[3]{\put(#1,#2){\line(0,1){#3}}}
\newcommand{\z}[3]{\put(#1,#2){\line(0,-1){#3}}}
\newcommand{\e}[3]{\put(#1,#2){\line(0,1){#3}}}
\y{   0}{   0}{   0}\x{   0}{   0}{ 133}
\e{  67}{    0}{   0}
\y{ 133}{   0}{   0}\x{ 133}{   0}{ 134}
\e{ 200}{    0}{   0}
\y{ 267}{   0}{  15}\x{ 267}{  15}{ 133}
\e{ 333}{   15}{   0}
\y{ 400}{  15}{ 119}\x{ 400}{ 134}{ 133}
\e{ 467}{  133}{   2}
\y{ 533}{ 134}{  16}\x{ 533}{ 150}{ 134}
\e{ 600}{  149}{   2}
\z{ 667}{ 150}{  44}\x{ 667}{ 106}{ 133}
\e{ 733}{  105}{   2}
\z{ 800}{ 106}{  57}\x{ 800}{  49}{ 133}
\e{ 867}{   48}{   2}
\z{ 933}{  49}{  37}\x{ 933}{  12}{ 134}
\e{1000}{   12}{   0}
\z{1067}{  12}{  11}\x{1067}{   1}{ 133}
\e{1133}{    1}{   0}
\end{picture}} 
\put(300,200){\begin{picture}( 1200,1200)
\thicklines 
\newcommand{\x}[3]{\put(#1,#2){\line(1,0){#3}}}
\newcommand{\y}[3]{\put(#1,#2){\line(0,1){#3}}}
\newcommand{\z}[3]{\put(#1,#2){\line(0,-1){#3}}}
\newcommand{\e}[3]{\put(#1,#2){\line(0,1){#3}}}
\y{   0}{   0}{   0}\x{   0}{   0}{ 133}
\e{  67}{    0}{   0}
\y{ 133}{   0}{   0}\x{ 133}{   0}{ 134}
\e{ 200}{    0}{   0}
\y{ 267}{   0}{  16}\x{ 267}{  16}{ 133}
\e{ 333}{   16}{   0}
\y{ 400}{  16}{ 132}\x{ 400}{ 148}{ 133}
\e{ 467}{  147}{   2}
\y{ 533}{ 148}{  41}\x{ 533}{ 189}{ 134}
\e{ 600}{  188}{   2}
\y{ 667}{ 189}{ 373}\x{ 667}{ 562}{ 133}
\e{ 733}{  547}{  28}
\z{ 800}{ 562}{ 518}\x{ 800}{  44}{ 133}
\e{ 867}{   44}{   2}
\z{ 933}{  44}{  33}\x{ 933}{  11}{ 134}
\e{1000}{   11}{   0}
\z{1067}{  11}{  10}\x{1067}{   1}{ 133}
\e{1133}{    1}{   0}
\end{picture}} 
\put(300,200){\begin{picture}( 1200,1200)
\newcommand{\R}[2]{\put(#1,#2){\circle*{ 12}}}
\newcommand{\E}[3]{\put(#1,#2){\line(0,1){#3}}}
\multiput(  0,  0)(25,0){11}{\circle*{12} }
\R{  67}{   0}
\E{  67}{    0}{   0}
\R{ 200}{   0}
\E{ 200}{    0}{   0}
\multiput(267,  8)(0,25){1}{\circle*{12} }
\multiput(267, 15)(25,0){6}{\circle*{12} }
\R{ 333}{  15}
\E{ 333}{   15}{   0}
\multiput(400, 15)(0,25){6}{\circle*{12} }
\multiput(400,138)(25,0){6}{\circle*{12} }
\R{ 467}{ 138}
\E{ 467}{  137}{   2}
\multiput(533,138)(0,25){3}{\circle*{12} }
\multiput(533,180)(25,0){6}{\circle*{12} }
\R{ 600}{ 180}
\E{ 600}{  179}{   2}
\multiput(667,180)(0,25){17}{\circle*{12} }
\multiput(667,582)(25,0){6}{\circle*{12} }
\R{ 733}{ 582}
\E{ 733}{  571}{  22}
\multiput(800, 41)(0,25){22}{\circle*{12} }
\multiput(800, 41)(25,0){6}{\circle*{12} }
\R{ 867}{  41}
\E{ 867}{   40}{   2}
\multiput(933, 11)(0,25){2}{\circle*{12} }
\multiput(933, 11)(25,0){6}{\circle*{12} }
\R{1000}{  11}
\E{1000}{   10}{   0}
\multiput(933,  1)(25,0){6}{\circle*{12} }
\R{1133}{   1}
\E{1133}{    1}{   0}
\end{picture}} 

\put(1900,200){\begin{picture}( 1200,1200)
\put(0,0){\framebox( 1200,1200){ }}
\multiput(  150.00,0)(  300.00,0){   4}{\line(0,1){25}}
\multiput(     .00,0)(   30.00,0){  41}{\line(0,1){10}}
\multiput(  150.00,1200)(  300.00,0){   4}{\line(0,-1){25}}
\multiput(     .00,1200)(   30.00,0){  41}{\line(0,-1){10}}
\put( 150,-25){\makebox(0,0)[t]{$    0.2 $}}
\put( 450,-25){\makebox(0,0)[t]{$    0.4 $}}
\put( 750,-25){\makebox(0,0)[t]{$    0.6 $}}
\put(1050,-25){\makebox(0,0)[t]{$    0.8 $}}
\put(1180,-100){\makebox(0,0)[t]{\large $    y_e $}}
\multiput(0,     .00)(0,  300.00){   5}{\line(1,0){25}}
\multiput(0,   30.00)(0,   30.00){  40}{\line(1,0){10}}
\multiput(1200,     .00)(0,  300.00){   5}{\line(-1,0){25}}
\multiput(1200,   30.00)(0,   30.00){  40}{\line(-1,0){10}}
\put(-25,   0){\makebox(0,0)[r]{$    0.0$}}
\put(-25, 300){\makebox(0,0)[r]{$    2.5$}}
\put(-25, 600){\makebox(0,0)[r]{$    5.0$}}
\put(-25, 900){\makebox(0,0)[r]{$    7.5$}}
\put(-25,1150){\makebox(0,0)[r]
              {{\small $\Delta y_e$}{\large $\frac{dN}{dy_e}$}}}
\put(600,1270){\makebox(0,0)[c]{\large\bf (d)} }
\put(100,1100){\makebox(0,0)[l]{\small $Q_e^2 > 15000\,GeV^2$}}
\end{picture}}
\put(1900,200){\begin{picture}( 1200,1200)
\newcommand{\R}[2]{\put(#1,#2){\makebox(0,0){$\star$}}}
\newcommand{\E}[3]{\put(#1,#2){\line(0,1){#3}}}
\R{ 450}{ 240}
\E{ 450}{   70}{ 340}
\R{ 750}{ 480}
\E{ 750}{  240}{ 480}
\R{1050}{ 720}
\E{1050}{  426}{ 588}
\end{picture}} 
\put(1900,200){\begin{picture}( 1200,1200)
\thinlines 
\newcommand{\x}[3]{\put(#1,#2){\line(1,0){#3}}}
\newcommand{\y}[3]{\put(#1,#2){\line(0,1){#3}}}
\newcommand{\z}[3]{\put(#1,#2){\line(0,-1){#3}}}
\newcommand{\e}[3]{\put(#1,#2){\line(0,1){#3}}}
\y{   0}{   0}{   7}\x{   0}{   7}{ 300}
\e{ 150}{    7}{   0}
\y{ 300}{   7}{ 117}\x{ 300}{ 124}{ 300}
\e{ 450}{  123}{   2}
\y{ 600}{ 124}{  69}\x{ 600}{ 193}{ 300}
\e{ 750}{  191}{   2}
\z{ 900}{ 193}{   3}\x{ 900}{ 190}{ 300}
\e{1050}{  188}{   2}
\end{picture}} 
\put(1900,200){\begin{picture}( 1200,1200)
\thicklines 
\newcommand{\x}[3]{\put(#1,#2){\line(1,0){#3}}}
\newcommand{\y}[3]{\put(#1,#2){\line(0,1){#3}}}
\newcommand{\z}[3]{\put(#1,#2){\line(0,-1){#3}}}
\newcommand{\e}[3]{\put(#1,#2){\line(0,1){#3}}}
\y{   0}{   0}{   7}\x{   0}{   7}{ 300}
\e{ 150}{    7}{   0}
\y{ 300}{   7}{ 235}\x{ 300}{ 242}{ 300}
\e{ 450}{  238}{   8}
\y{ 600}{ 242}{ 168}\x{ 600}{ 410}{ 300}
\e{ 750}{  402}{  16}
\y{ 900}{ 410}{   1}\x{ 900}{ 411}{ 300}
\e{1050}{  398}{  26}
\end{picture}} 
\put(1900,200){\begin{picture}( 1200,1200)
\newcommand{\R}[2]{\put(#1,#2){\circle*{ 12}}}
\newcommand{\E}[3]{\put(#1,#2){\line(0,1){#3}}}
\multiput(  0,  7)(25,0){13}{\circle*{12} }
\multiput(300,  7)(0,25){15}{\circle*{12} }
\R{ 150}{   7}
\E{ 150}{    6}{   0}
\multiput(600,376)(0,25){2}{\circle*{12} }
\multiput(300,376)(25,0){13}{\circle*{12} }
\R{ 450}{ 376}
\E{ 450}{  368}{  16}
\multiput(600,428)(25,0){13}{\circle*{12} }
\R{ 750}{ 428}
\E{ 750}{  420}{  16}
\multiput(900,254)(0,25){8}{\circle*{12} }
\multiput(900,254)(25,0){13}{\circle*{12} }
\R{1050}{ 254}
\E{1050}{  250}{   8}
\end{picture}} 
 
\end{picture} 
\caption[fig2]{\small\sf 
Distributions of $Q_e^2$, $y_e$ and $M_e$ obtained from LESKO-YFS for the
Standard Model process (thin-line histograms) and in the presence
of the scalar (thick-line histograms) and vector (dotted histograms)
leptoquarks in models (1) and (2) with 
\protect$M_X=200\,GeV,\;\Gamma_X=2\,GeV,\;\delta_S=-0.72,\,\delta_V=-0.68$ 
(see the text for details) 
compared with the recent H1 high $Q^2$ events ($\star$ symbols) as given 
in Ref.~\cite{h1}. 
\protect$\Delta v\,(v=Q_e^2,y_e,M_e)$ denotes bin size for respective
histograms. 
}
\label{fig2}

\end{figure} 


\begin{figure}[!ht]
\centering
\setlength{\unitlength}{0.05mm}
\begin{picture}(3200,1550)

\put(250,200){\begin{picture}( 1200,1200)
\put(0,0){\framebox( 1200,1200){ }}
\multiput(  300.00,0)(  300.00,0){   4}{\line(0,1){25}}
\multiput(     .00,0)(   30.00,0){  41}{\line(0,1){10}}
\multiput(  300.00,1200)(  300.00,0){   4}{\line(0,-1){25}}
\multiput(     .00,1200)(   30.00,0){  41}{\line(0,-1){10}}
\put(   0,-25){\makebox(0,0)[t]{$   0.0 $}}
\put( 300,-25){\makebox(0,0)[t]{$   0.25 $}}
\put( 600,-25){\makebox(0,0)[t]{$   0.5 $}}
\put( 900,-25){\makebox(0,0)[t]{$   0.75 $}}
\put(1200,-25){\makebox(0,0)[t]{$   1.0 $}}
\put(1050,-100){\makebox(0,0)[t]{\large $ x_{DA} $}}
\multiput(0,     .00)(0,  240.00){   6}{\line(1,0){25}}
\multiput(0,   72.24)(0,  240.00){   5}{\line(1,0){10}}
\multiput(0,  114.04)(0,  240.00){   5}{\line(1,0){10}}
\multiput(0,  144.50)(0,  240.00){   5}{\line(1,0){10}}
\multiput(0,  167.76)(0,  240.00){   5}{\line(1,0){10}}
\multiput(0,  186.77)(0,  240.00){   5}{\line(1,0){10}}
\multiput(0,  202.82)(0,  240.00){   5}{\line(1,0){10}}
\multiput(0,  216.74)(0,  240.00){   5}{\line(1,0){10}}
\multiput(0,  229.01)(0,  240.00){   5}{\line(1,0){10}}
\multiput(1200,     .00)(0,  240.00){   6}{\line(-1,0){25}}
\multiput(1200,   72.24)(0,  240.00){   5}{\line(-1,0){10}}
\multiput(1200,  114.04)(0,  240.00){   5}{\line(-1,0){10}}
\multiput(1200,  144.50)(0,  240.00){   5}{\line(-1,0){10}}
\multiput(1200,  167.76)(0,  240.00){   5}{\line(-1,0){10}}
\multiput(1200,  186.77)(0,  240.00){   5}{\line(-1,0){10}}
\multiput(1200,  202.82)(0,  240.00){   5}{\line(-1,0){10}}
\multiput(1200,  216.74)(0,  240.00){   5}{\line(-1,0){10}}
\multiput(1200,  229.01)(0,  240.00){   5}{\line(-1,0){10}}
\put(-25,   0){\makebox(0,0)[r]{$   10^{-3} $}}
\put(-25, 240){\makebox(0,0)[r]{$   10^{-2} $}}
\put(-25, 480){\makebox(0,0)[r]{$   10^{-1} $}}
\put(-25, 720){\makebox(0,0)[r]{$     1 $}}
\put(-25, 960){\makebox(0,0)[r]{$    10 $}}
\put(-25,1150){\makebox(0,0)[r]{
              {\small $\Delta x_{DA}$}{\large $\frac{dN}{dx_{DA}}$}}}
\put(600,1270){\makebox(0,0)[c]{\large\bf (a)} }
\put(550,1130){\makebox(0,0)[l]{\small $y_{DA} > 0.25$}}
\put(550,1050){\makebox(0,0)[l]{\small $Q_{DA}^2 > 5000\,GeV^2$}}
\end{picture}}
\put(250,200){\begin{picture}( 1200,1200)
\newcommand{\R}[2]{\put(#1,#2){\makebox(0,0){$\diamond$}}}
\newcommand{\E}[3]{\put(#1,#2){\line(0,1){#3}}}
\R{  90}{1060}
\E{  90}{ 1037}{  41}
\R{ 150}{1075}
\E{ 150}{ 1053}{  39}
\R{ 210}{1104}
\E{ 210}{ 1087}{  33}
\R{ 270}{1060}
\E{ 270}{ 1037}{  41}
\R{ 330}{ 987}
\E{ 330}{  954}{  59}
\R{ 390}{ 864}
\E{ 390}{  792}{ 114}
\R{ 450}{ 864}
\E{ 450}{  792}{ 114}
\R{ 510}{ 864}
\E{ 510}{  792}{ 114}
\R{ 570}{ 792}
\E{ 570}{  664}{ 184}
\R{ 690}{ 792}
\E{ 690}{  664}{ 184}
\R{ 750}{ 720}
\E{ 750}{    0}{ 792}
\R{ 870}{ 720}
\E{ 870}{    0}{ 792}
\end{picture}} 
\put(250,200){\begin{picture}( 1200,1200)
\thinlines 
\newcommand{\x}[3]{\put(#1,#2){\line(1,0){#3}}}
\newcommand{\y}[3]{\put(#1,#2){\line(0,1){#3}}}
\newcommand{\z}[3]{\put(#1,#2){\line(0,-1){#3}}}
\newcommand{\e}[3]{\put(#1,#2){\line(0,1){#3}}}
\y{   0}{   0}{   0}\x{   0}{   0}{  60}
\e{  30}{    0}{   0}
\y{  60}{   0}{1081}\x{  60}{1081}{  60}
\e{  90}{ 1081}{   0}
\y{ 120}{1081}{  34}\x{ 120}{1115}{  60}
\e{ 150}{ 1115}{   0}
\z{ 180}{1115}{  24}\x{ 180}{1091}{  60}
\e{ 210}{ 1090}{   0}
\z{ 240}{1091}{  35}\x{ 240}{1056}{  60}
\e{ 270}{ 1056}{   0}
\z{ 300}{1056}{  62}\x{ 300}{ 994}{  60}
\e{ 330}{  993}{   0}
\z{ 360}{ 994}{  63}\x{ 360}{ 931}{  60}
\e{ 390}{  930}{   0}
\z{ 420}{ 931}{  59}\x{ 420}{ 872}{  60}
\e{ 450}{  871}{   2}
\z{ 480}{ 872}{  59}\x{ 480}{ 813}{  60}
\e{ 510}{  812}{   2}
\z{ 540}{ 813}{  58}\x{ 540}{ 755}{  60}
\e{ 570}{  754}{   2}
\z{ 600}{ 755}{  63}\x{ 600}{ 692}{  60}
\e{ 630}{  691}{   2}
\z{ 660}{ 692}{  63}\x{ 660}{ 629}{  60}
\e{ 690}{  627}{   4}
\z{ 720}{ 629}{  75}\x{ 720}{ 554}{  60}
\e{ 750}{  551}{   6}
\z{ 780}{ 554}{  81}\x{ 780}{ 473}{  60}
\e{ 810}{  469}{   8}
\z{ 840}{ 473}{  91}\x{ 840}{ 382}{  60}
\e{ 870}{  376}{  12}
\z{ 900}{ 382}{  89}\x{ 900}{ 293}{  60}
\e{ 930}{  283}{  20}
\z{ 960}{ 293}{ 109}\x{ 960}{ 184}{  60}
\e{ 990}{  168}{  30}
\z{1020}{ 184}{ 126}\x{1020}{  58}{  60}
\e{1050}{   25}{  50}
\z{1080}{  58}{  58}\x{1080}{   0}{  60}
\e{1110}{    0}{   0}
\y{1140}{   0}{   0}\x{1140}{   0}{  60}
\e{1170}{    0}{   0}
\end{picture}} 
\put(250,200){\begin{picture}( 1200,1200)
\thicklines 
\newcommand{\x}[3]{\put(#1,#2){\line(1,0){#3}}}
\newcommand{\y}[3]{\put(#1,#2){\line(0,1){#3}}}
\newcommand{\z}[3]{\put(#1,#2){\line(0,-1){#3}}}
\newcommand{\e}[3]{\put(#1,#2){\line(0,1){#3}}}
\y{   0}{   0}{   0}\x{   0}{   0}{  60}
\e{  30}{    0}{   0}
\y{  60}{   0}{1082}\x{  60}{1082}{  60}
\e{  90}{ 1082}{   0}
\y{ 120}{1082}{  34}\x{ 120}{1116}{  60}
\e{ 150}{ 1116}{   0}
\z{ 180}{1116}{  23}\x{ 180}{1093}{  60}
\e{ 210}{ 1092}{   0}
\z{ 240}{1093}{  34}\x{ 240}{1059}{  60}
\e{ 270}{ 1059}{   0}
\z{ 300}{1059}{  61}\x{ 300}{ 998}{  60}
\e{ 330}{  998}{   0}
\z{ 360}{ 998}{  59}\x{ 360}{ 939}{  60}
\e{ 390}{  939}{   0}
\z{ 420}{ 939}{  47}\x{ 420}{ 892}{  60}
\e{ 450}{  891}{   2}
\y{ 480}{ 892}{  78}\x{ 480}{ 970}{  60}
\e{ 510}{  967}{   6}
\z{ 540}{ 970}{ 138}\x{ 540}{ 832}{  60}
\e{ 570}{  829}{   6}
\z{ 600}{ 832}{ 119}\x{ 600}{ 713}{  60}
\e{ 630}{  708}{  10}
\z{ 660}{ 713}{  93}\x{ 660}{ 620}{  60}
\e{ 690}{  619}{   4}
\z{ 720}{ 620}{  74}\x{ 720}{ 546}{  60}
\e{ 750}{  543}{   6}
\z{ 780}{ 546}{  80}\x{ 780}{ 466}{  60}
\e{ 810}{  462}{   8}
\z{ 840}{ 466}{  91}\x{ 840}{ 375}{  60}
\e{ 870}{  369}{  12}
\z{ 900}{ 375}{  88}\x{ 900}{ 287}{  60}
\e{ 930}{  277}{  20}
\z{ 960}{ 287}{ 109}\x{ 960}{ 178}{  60}
\e{ 990}{  162}{  30}
\z{1020}{ 178}{ 128}\x{1020}{  50}{  60}
\e{1050}{   17}{  50}
\z{1080}{  50}{  50}\x{1080}{   0}{  60}
\e{1110}{    0}{   0}
\y{1140}{   0}{   0}\x{1140}{   0}{  60}
\e{1170}{    0}{   0}
\end{picture}} 
\put(250,200){\begin{picture}( 1200,1200)
\newcommand{\R}[2]{\put(#1,#2){\circle*{ 12}}}
\newcommand{\E}[3]{\put(#1,#2){\line(0,1){#3}}}
\multiput(  0,  0)(30,0){3}{\circle*{12} }
\multiput( 60,  0)(0,25){44}{\circle*{12} }
\R{  30}{   0}
\E{  30}{    0}{   0}
\multiput( 60,1081)(30,0){3}{\circle*{12} }
\multiput(120,1081)(0,25){2}{\circle*{12} }
\R{  90}{1081}
\E{  90}{ 1081}{   0}
\multiput(120,1116)(30,0){3}{\circle*{12} }
\multiput(180,1092)(0,25){2}{\circle*{12} }
\R{ 150}{1116}
\E{ 150}{ 1115}{   0}
\multiput(180,1092)(30,0){3}{\circle*{12} }
\multiput(240,1059)(0,25){2}{\circle*{12} }
\R{ 210}{1092}
\E{ 210}{ 1092}{   0}
\multiput(240,1059)(30,0){3}{\circle*{12} }
\multiput(300, 999)(0,25){3}{\circle*{12} }
\R{ 270}{1059}
\E{ 270}{ 1059}{   0}
\multiput(300, 999)(30,0){3}{\circle*{12} }
\multiput(360, 941)(0,25){3}{\circle*{12} }
\R{ 330}{ 999}
\E{ 330}{  999}{   0}
\multiput(360, 941)(30,0){3}{\circle*{12} }
\multiput(420, 897)(0,25){3}{\circle*{12} }
\R{ 390}{ 941}
\E{ 390}{  941}{   0}
\multiput(420, 897)(30,0){3}{\circle*{12} }
\multiput(480, 897)(0,25){4}{\circle*{12} }
\R{ 450}{ 897}
\E{ 450}{  897}{   2}
\multiput(480, 993)(30,0){3}{\circle*{12} }
\multiput(540, 846)(0,25){7}{\circle*{12} }
\R{ 510}{ 993}
\E{ 510}{  992}{   2}
\multiput(540, 846)(30,0){3}{\circle*{12} }
\multiput(600, 707)(0,25){6}{\circle*{12} }
\R{ 570}{ 846}
\E{ 570}{  843}{   4}
\multiput(600, 707)(30,0){3}{\circle*{12} }
\multiput(660, 612)(0,25){4}{\circle*{12} }
\R{ 630}{ 707}
\E{ 630}{  703}{   6}
\multiput(660, 612)(30,0){3}{\circle*{12} }
\multiput(720, 538)(0,25){4}{\circle*{12} }
\R{ 690}{ 612}
\E{ 690}{  610}{   4}
\multiput(720, 538)(30,0){3}{\circle*{12} }
\multiput(780, 459)(0,25){4}{\circle*{12} }
\R{ 750}{ 538}
\E{ 750}{  536}{   6}
\multiput(780, 459)(30,0){3}{\circle*{12} }
\multiput(840, 369)(0,25){4}{\circle*{12} }
\R{ 810}{ 459}
\E{ 810}{  455}{   8}
\multiput(840, 369)(30,0){3}{\circle*{12} }
\multiput(900, 281)(0,25){4}{\circle*{12} }
\R{ 870}{ 369}
\E{ 870}{  363}{  12}
\multiput(900, 281)(30,0){3}{\circle*{12} }
\multiput(960, 174)(0,25){5}{\circle*{12} }
\R{ 930}{ 281}
\E{ 930}{  271}{  20}
\multiput( 960, 174)(30,0){3}{\circle*{12} }
\multiput(1020,  45)(0,25){6}{\circle*{12} }
\R{ 990}{ 174}
\E{ 990}{  157}{  30}
\multiput(1020,  45)(30,0){3}{\circle*{12} }
\multiput(1080,   0)(0,25){2}{\circle*{12} }
\R{1050}{  45}
\E{1050}{   13}{  50}
\multiput(1080,  0)(30,0){5}{\circle*{12} }
\R{1110}{   0}
\E{1110}{    0}{   0}
\R{1170}{   0}
\E{1170}{    0}{   0}
\end{picture}} 

\put(1950,200){\begin{picture}( 1200,1200)
\put(0,0){\framebox( 1200,1200){ }}
\multiput(  133.33,0)(  266.67,0){   5}{\line(0,1){25}}
\multiput(     .00,0)(   26.67,0){  46}{\line(0,1){10}}
\multiput(  133.33,1200)(  266.67,0){   5}{\line(0,-1){25}}
\multiput(     .00,1200)(   26.67,0){  46}{\line(0,-1){10}}
\put( 133,-25){\makebox(0,0)[t]{$    10^{   4} $}}
\put( 400,-25){\makebox(0,0)[t]{$    2\cdot 10^{   4} $}}
\put( 667,-25){\makebox(0,0)[t]{$    3\cdot 10^{   4} $}}
\put( 933,-25){\makebox(0,0)[t]{$    4\cdot 10^{   4} $}}
\put(1200,-25){\makebox(0,0)[t]{$    5\cdot 10^{   4} $}}
\put(1100,-120){\makebox(0,0)[t]{$Q_{DA}^2 [GeV^2] $}}
\multiput(0,     .00)(0,  200.00){   7}{\line(1,0){25}}
\multiput(0,   60.20)(0,  200.00){   6}{\line(1,0){10}}
\multiput(0,   95.42)(0,  200.00){   6}{\line(1,0){10}}
\multiput(0,  120.42)(0,  200.00){   6}{\line(1,0){10}}
\multiput(0,  139.80)(0,  200.00){   6}{\line(1,0){10}}
\multiput(0,  155.64)(0,  200.00){   6}{\line(1,0){10}}
\multiput(0,  169.02)(0,  200.00){   6}{\line(1,0){10}}
\multiput(0,  180.62)(0,  200.00){   6}{\line(1,0){10}}
\multiput(0,  190.84)(0,  200.00){   6}{\line(1,0){10}}
\multiput(1200,     .00)(0,  200.00){   7}{\line(-1,0){25}}
\multiput(1200,   60.20)(0,  200.00){   6}{\line(-1,0){10}}
\multiput(1200,   95.42)(0,  200.00){   6}{\line(-1,0){10}}
\multiput(1200,  120.42)(0,  200.00){   6}{\line(-1,0){10}}
\multiput(1200,  139.80)(0,  200.00){   6}{\line(-1,0){10}}
\multiput(1200,  155.64)(0,  200.00){   6}{\line(-1,0){10}}
\multiput(1200,  169.02)(0,  200.00){   6}{\line(-1,0){10}}
\multiput(1200,  180.62)(0,  200.00){   6}{\line(-1,0){10}}
\multiput(1200,  190.84)(0,  200.00){   6}{\line(-1,0){10}}
\put(-25,   0){\makebox(0,0)[r]{$   10^{-3} $}}
\put(-25, 200){\makebox(0,0)[r]{$   10^{-2} $}}
\put(-25, 400){\makebox(0,0)[r]{$   10^{-1} $}}
\put(-25, 600){\makebox(0,0)[r]{$     1 $}}
\put(-25, 800){\makebox(0,0)[r]{$    10 $}}
\put(-25,1000){\makebox(0,0)[r]{$    10^2 $}}
\put(-25,1150){\makebox(0,0)[r]
              {{\small $\Delta Q_{DA}^2$}{\large $\frac{dN}{dQ_{DA}^2}$}}}
\put(600,1270){\makebox(0,0)[c]{\large\bf (b)} }
\thinlines
\put(450,1100){\line(1,0){70} }
\put(550,1080){\small SM}
\thicklines
\put(450,1020){\line(1,0){70} }
\put(550,1000){\small Scalar LQ}
\multiput(460,940)(25,0){3}{\circle*{12} }
\put(550, 920){\small Vector LQ}
\put(470, 840){$\diamond$}
\put(550, 840){\small ZEUS data points}
\thinlines
\end{picture}}
\put(1950,200){\begin{picture}( 1200,1200)
\newcommand{\R}[2]{\put(#1,#2){\makebox(0,0){$\diamond$}}}
\newcommand{\E}[3]{\put(#1,#2){\line(0,1){#3}}}
\R{  33}{1018}
\E{  33}{ 1010}{  15}
\R{ 100}{ 909}
\E{ 100}{  892}{  31}
\R{ 167}{ 841}
\E{ 167}{  816}{  44}
\R{ 233}{ 740}
\E{ 233}{  688}{  83}
\R{ 300}{ 756}
\E{ 300}{  710}{  76}
\R{ 367}{ 600}
\E{ 367}{    0}{ 660}
\R{ 433}{ 660}
\E{ 433}{  554}{ 154}
\R{ 567}{ 600}
\E{ 567}{    0}{ 660}
\R{ 833}{ 600}
\E{ 833}{    0}{ 660}
\R{1100}{ 600}
\E{1100}{    0}{ 660}
\end{picture}} 
\put(1950,200){\begin{picture}( 1200,1200)
\thinlines 
\newcommand{\x}[3]{\put(#1,#2){\line(1,0){#3}}}
\newcommand{\y}[3]{\put(#1,#2){\line(0,1){#3}}}
\newcommand{\z}[3]{\put(#1,#2){\line(0,-1){#3}}}
\newcommand{\e}[3]{\put(#1,#2){\line(0,1){#3}}}
\y{   0}{   0}{1021}\x{   0}{1021}{  67}
\e{  33}{ 1021}{   0}
\z{  67}{1021}{  98}\x{  67}{ 923}{  66}
\e{ 100}{  923}{   0}
\z{ 133}{ 923}{  77}\x{ 133}{ 846}{  67}
\e{ 167}{  845}{   0}
\z{ 200}{ 846}{  67}\x{ 200}{ 779}{  67}
\e{ 233}{  779}{   0}
\z{ 267}{ 779}{  59}\x{ 267}{ 720}{  66}
\e{ 300}{  720}{   2}
\z{ 333}{ 720}{  54}\x{ 333}{ 666}{  67}
\e{ 367}{  665}{   2}
\z{ 400}{ 666}{  52}\x{ 400}{ 614}{  67}
\e{ 433}{  613}{   2}
\z{ 467}{ 614}{  44}\x{ 467}{ 570}{  66}
\e{ 500}{  569}{   2}
\z{ 533}{ 570}{  45}\x{ 533}{ 525}{  67}
\e{ 567}{  523}{   4}
\z{ 600}{ 525}{  44}\x{ 600}{ 481}{  67}
\e{ 633}{  479}{   4}
\z{ 667}{ 481}{  41}\x{ 667}{ 440}{  66}
\e{ 700}{  437}{   4}
\z{ 733}{ 440}{  44}\x{ 733}{ 396}{  67}
\e{ 767}{  393}{   6}
\z{ 800}{ 396}{  43}\x{ 800}{ 353}{  67}
\e{ 833}{  349}{   8}
\z{ 867}{ 353}{  41}\x{ 867}{ 312}{  66}
\e{ 900}{  307}{  10}
\z{ 933}{ 312}{  35}\x{ 933}{ 277}{  67}
\e{ 967}{  271}{  12}
\z{1000}{ 277}{  33}\x{1000}{ 244}{  67}
\e{1033}{  236}{  16}
\z{1067}{ 244}{  40}\x{1067}{ 204}{  66}
\e{1100}{  194}{  20}
\z{1133}{ 204}{  45}\x{1133}{ 159}{  67}
\e{1167}{  146}{  26}
\end{picture}} 
\put(1950,200){\begin{picture}( 1200,1200)
\thicklines 
\newcommand{\x}[3]{\put(#1,#2){\line(1,0){#3}}}
\newcommand{\y}[3]{\put(#1,#2){\line(0,1){#3}}}
\newcommand{\z}[3]{\put(#1,#2){\line(0,-1){#3}}}
\newcommand{\e}[3]{\put(#1,#2){\line(0,1){#3}}}
\y{   0}{   0}{1022}\x{   0}{1022}{  67}
\e{  33}{ 1022}{   0}
\z{  67}{1022}{  95}\x{  67}{ 927}{  66}
\e{ 100}{  927}{   0}
\z{ 133}{ 927}{  74}\x{ 133}{ 853}{  67}
\e{ 167}{  853}{   0}
\z{ 200}{ 853}{  60}\x{ 200}{ 793}{  67}
\e{ 233}{  792}{   0}
\z{ 267}{ 793}{  49}\x{ 267}{ 744}{  66}
\e{ 300}{  743}{   2}
\z{ 333}{ 744}{  42}\x{ 333}{ 702}{  67}
\e{ 367}{  701}{   2}
\z{ 400}{ 702}{  34}\x{ 400}{ 668}{  67}
\e{ 433}{  666}{   4}
\z{ 467}{ 668}{  21}\x{ 467}{ 647}{  66}
\e{ 500}{  643}{   6}
\z{ 533}{ 647}{  14}\x{ 533}{ 633}{  67}
\e{ 567}{  629}{   8}
\z{ 600}{ 633}{  25}\x{ 600}{ 608}{  67}
\e{ 633}{  602}{  12}
\y{ 667}{ 608}{   5}\x{ 667}{ 613}{  66}
\e{ 700}{  603}{  18}
\z{ 733}{ 613}{  18}\x{ 733}{ 595}{  67}
\e{ 767}{  583}{  21}
\z{ 800}{ 595}{   7}\x{ 800}{ 588}{  67}
\e{ 833}{  572}{  35}
\z{ 867}{ 588}{  24}\x{ 867}{ 564}{  66}
\e{ 900}{  546}{  32}
\z{ 933}{ 564}{ 151}\x{ 933}{ 413}{  67}
\e{ 967}{  383}{  50}
\z{1000}{ 413}{  81}\x{1000}{ 332}{  67}
\e{1033}{  290}{  70}
\z{1067}{ 332}{  13}\x{1067}{ 319}{  66}
\e{1100}{  246}{ 110}
\z{1133}{ 319}{ 183}\x{1133}{ 136}{  67}
\e{1167}{  123}{  22}
\end{picture}} 
\put(1950,200){\begin{picture}( 1200,1200)
\newcommand{\R}[2]{\put(#1,#2){\circle*{ 12}}}
\newcommand{\E}[3]{\put(#1,#2){\line(0,1){#3}}}
\multiput(  0,1025)(32,0){3}{\circle*{12} }
\multiput( 67, 933)(0,25){4}{\circle*{12} }
\R{  33}{1025}
\E{  33}{ 1025}{   0}
\multiput( 67, 933)(32,0){3}{\circle*{12} }
\multiput(133, 863)(0,25){3}{\circle*{12} }
\R{ 100}{ 933}
\E{ 100}{  933}{   0}
\multiput(133, 863)(32,0){3}{\circle*{12} }
\multiput(200, 806)(0,25){3}{\circle*{12} }
\R{ 167}{ 863}
\E{ 167}{  863}{   0}
\multiput(200, 806)(32,0){3}{\circle*{12} }
\multiput(267, 761)(0,25){2}{\circle*{12} }
\R{ 233}{ 807}
\E{ 233}{  806}{   2}
\multiput(267, 761)(32,0){3}{\circle*{12} }
\multiput(333, 719)(0,25){2}{\circle*{12} }
\R{ 300}{ 761}
\E{ 300}{  760}{   2}
\multiput(333, 719)(32,0){3}{\circle*{12} }
\multiput(400, 679)(0,25){2}{\circle*{12} }
\R{ 367}{ 719}
\E{ 367}{  717}{   4}
\multiput(400, 679)(32,0){3}{\circle*{12} }
\multiput(467, 647)(0,25){2}{\circle*{12} }
\R{ 433}{ 679}
\E{ 433}{  676}{   4}
\multiput(467, 647)(32,0){3}{\circle*{12} }
\multiput(533, 614)(0,25){2}{\circle*{12} }
\R{ 500}{ 647}
\E{ 500}{  643}{   8}
\multiput(533, 614)(32,0){3}{\circle*{12} }
\multiput(600, 566)(0,25){3}{\circle*{12} }
\R{ 567}{ 614}
\E{ 567}{  610}{   8}
\multiput(600, 566)(32,0){3}{\circle*{12} }
\multiput(667, 535)(0,25){2}{\circle*{12} }
\R{ 633}{ 566}
\E{ 633}{  561}{  10}
\multiput(667, 535)(32,0){3}{\circle*{12} }
\multiput(733, 476)(0,25){3}{\circle*{12} }
\R{ 700}{ 535}
\E{ 700}{  527}{  16}
\multiput(733, 476)(32,0){3}{\circle*{12} }
\multiput(800, 408)(0,25){3}{\circle*{12} }
\R{ 767}{ 476}
\E{ 767}{  467}{  18}
\multiput(800, 408)(32,0){3}{\circle*{12} }
\multiput(867, 340)(0,25){3}{\circle*{12} }
\R{ 833}{ 408}
\E{ 833}{  400}{  18}
\multiput(867, 340)(32,0){3}{\circle*{12} }
\multiput(933, 278)(0,25){3}{\circle*{12} }
\R{ 900}{ 340}
\E{ 900}{  329}{  20}
\multiput( 933, 278)(32,0){3}{\circle*{12} }
\multiput(1000, 249)(0,25){2}{\circle*{12} }
\R{ 967}{ 278}
\E{ 967}{  271}{  14}
\multiput(1000, 249)(32,0){3}{\circle*{12} }
\multiput(1067, 201)(0,25){2}{\circle*{12} }
\R{1033}{ 249}
\E{1033}{  239}{  20}
\multiput(1067, 201)(32,0){3}{\circle*{12} }
\multiput(1133, 155)(0,25){3}{\circle*{12} }
\R{1100}{ 201}
\E{1100}{  191}{  20}
\multiput(1133, 155)(32,0){3}{\circle*{12} }
\R{1167}{ 155}
\E{1167}{  142}{  26}
\end{picture}} 

\end{picture} 
\caption[fig2]{\small\sf 
Distributions of $x_{DA}$ and $Q_{DA}^2$ obtained from LESKO-YFS for the
Standard Model process (thin-line histograms) and in the presence
of the same scalar (thick-line histograms) and vector (dotted histograms)
leptoquarks as in Fig.~\ref{fig2}
compared with the recent ZEUS high $Q^2$ events ($\diamond$ symbols) as given 
in Ref.~\cite{zeus}. 
\protect$\Delta v\,(v=x_{DA},Q_{DA}^2)$ denotes bin size for respective
histograms. 
}
\label{fig3}

\end{figure}
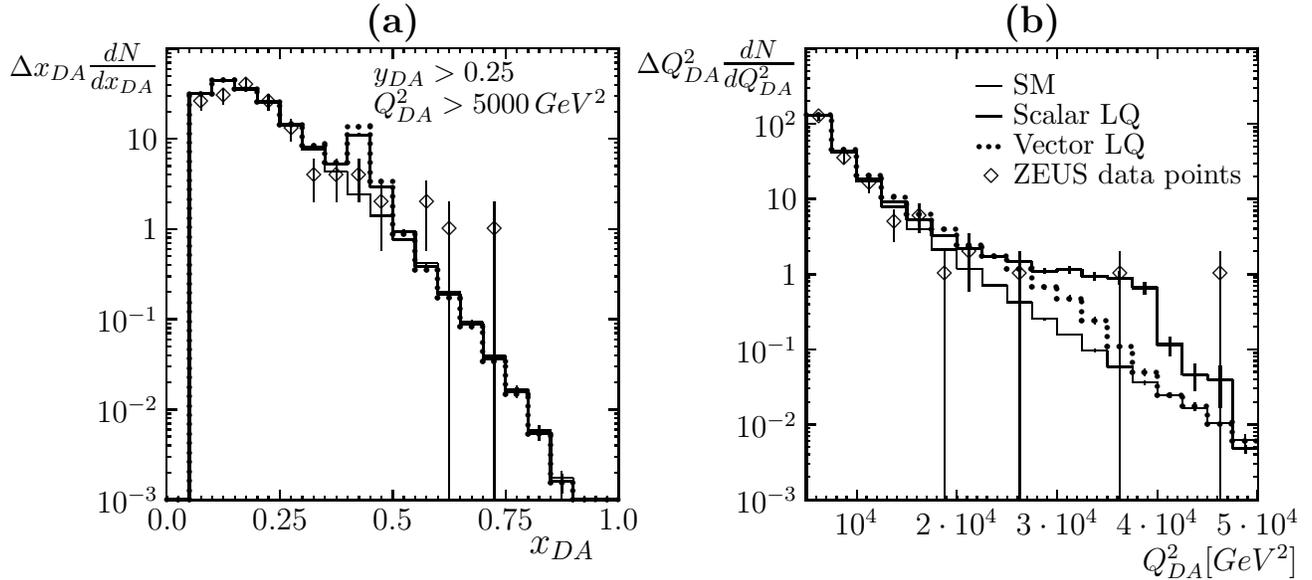 

We then show in Figs.~2 and 3 the comparison of the H1 and
ZEUS data, respectively,
with the models for the observables $Q_e^2$, $y_e$ and $M_e$
for H1, and the observables $x_{DA}$ and $Q_{DA}$ (see
Refs.~\cite{h1,zeus} for their definition) for ZEUS,
where for H1 we show data for 
$Q_e^2$ with $0.1<y_e<0.9$, for $y_e$ with $M_e> 180$ GeV,
for $M_e$ with $Q_e^2>15000$ GeV$^2$, $0.1<y_e<0.9$, and for
$y_e$ with $Q_e^2>15000$ GeV$^2$ and for ZEUS we show
data for $x_{DA}$ with $y_{DA}> 0.25$, $Q_{DA}^2>5000$ GeV$^2$, and for
$Q_{DA}^2$ in the regions of the anomalous numbers of events. 
The respective experimental acceptances~\cite{h1,zeus} are included in 
the above results.
In all six distributions, we see that, for
$\delta\cong -0.7$, there is reasonable agreement between the data
and the theoretical prediction for models (1) and (2) for $S=0$ whereas
for $S=1$ the comparison with the data in Figs.~2a,b,d and 3a,b is
less favorable but still tolerable (within $3\sigma$).
Thus, we must await more data for a detailed check of the models (1) and~(2).  
\par
In our plots we have used the upper limit~\cite{h1} value
$\Gamma_X= 2\,$GeV; if we use the theoretical lower limit
$\Gamma_X=g_X^2M_X/32\pi(48\pi)$ for $S=0(1)$ the required value of
$\delta$ in Figs. 2,~3 changes to $\sim -0.95$ but our 
qualitative conclusions do not change.
We have also checked that varying our
structure functions between those in Ref.~\cite{mftung:1991} and those in
Refs.~\cite{MRSA:1994} does not affect our results in any significant
way at the level of precision of the HERA data. What
we can do is to emphasize that, if models (1) and (2) are
indeed correct, we do not expect similar anomalies in the $e^-$
data because that would involve scattering from the sea and this
would be strongly suppressed. Further, we do not have to have
charged current events: to get them, we need to add more terms
to the interactions in the ${\cal L}^{\{Q,Q-1\}}_{int,S,H}$ given above
but in our general
phenomenological framework, this is neither mandated nor forbidden.
We also comment on the expectations at FNAL. What they are
in general depends on the assumed BR to the initial state channel
at HERA for the final state decay as well as on the detailed
coupling scenario between gluons and $X$:
as emphasized in Ref.~\cite{guido:1997}, 
a BR near unity with only two higher multipoles
in the gluon-$X$ coupling scenario
may already be disfavored by the
D0+CDF~\cite{cdf:1993,d02:1994,D0} limits if this combination can be done 
without unforseen systematics although as Ref.~\cite{blumlein:1997}
has stressed the situation depends in detail on cuts, BR's, etc. and 
the implied signal may be small enough that it is missed and hence the matter
is still unsettled;
but, in our general phenomenological
approach, we have no reason to expect a BR near unity or a
naive coupling between leptoquarks and gluons: an infinite series
of multipoles is in principle allowed in our effective low energy
Lagrangian for X and gluons, with the truly underlying renormalizable
theory awaiting yet higher energy probes to reveal itself.
Thus, we can only encourage the FNAL experiments to continue
their search for the observation of $X$ as well.
\par 
In summary, we have investigated the comparison with 
HERA data of the expectations
on the leptoquark interpretation of the anomalous positron-jet
phenomenon at DESY. We used the 
LESKO-YFS~\cite{lesko-yfs:1992,placzekPHD:1993} Monte Carlo event generator
so that higher radiative corrections to the Born level leptoquark
signal are calculated with the YFS exponentiated ${\cal O}(\alpha)$ LL
$\bar\beta_{0,1}$ residuals, for both initial and final state radiation with
finite-p$_T$ effects in this radiation properly realized,
in the framework of Ref.~\cite{bhlumi4:1996,bhlumi2:1992}.
We find that, in agreement with the analysis in Ref.~\cite{sjbwzw:1997},
a leptoquark coupling of $\lesssim 0.3g_W$ in our models (1) and (2),
positron-quark resonances, where spin 0 is preferred over spin 1,
is in general agreement with the HERA data distributions at both
high and low $Q^2$ for $\Gamma_X\lesssim 2$GeV
and that our models (3) and (4), positron-anti-quark
resonances, like the SM itself, even in the presence of the rigorous
higher order initial and final state radiation with
finite-p$_T$ photon effects,
cannot account for the data for
any reasonable value of the respective leptoquark coupling parameter.
We look forward with excitement
to more precise data.
\par
\vskip .5cm
{\bf \Large Acknowledgements}\\

Two of us (S.J. and B.F.L.W.) acknowledge 
the kind hospitality of Prof. G. Veneziano
and the CERN TH Division and the support of 
Prof. D. Schlatter and the ALEPH
Collaboration while this work was done. We thank Dr. S.-M. Kim for
assistance with the implementation of the MRSA structure functions.

\bibliographystyle{prsty}
\bibliography{97-0103p}

\end{document}